       \documentclass[twocolumn,prb,aps,a4paper,epsf,showkeys,superscriptaddress,longbibliography] {revtex4} 

\usepackage{etoolbox}
\usepackage{fancyhdr} 
\newtoggle{usualpreprint} \newtoggle{journal-like} \newtoggle{journal-like-bigfigures}

  \togglefalse{usualpreprint} \toggletrue{journal-like}  \togglefalse{journal-like-bigfigures} 

\newcommand{\myheading}{IVCT luminescence in Cs$_2$LiLuCl$_6$:Ce}
\newcommand{\Authors}{Seijo and Barandiar\'an}
\newcommand{\FIGa}{CeIII}

\newcommand{\FIGc}{CeIII-CeIV-groundstate}
\newcommand{\FIGd}{CeIII-CeIV-several}

\newcommand{\FIGf}{ivctl}
\newcommand{\FIGg}{ivct}
\newcommand{\FIGh}{gs_surfaces}
\newcommand{\FIGi}{Qet}
\newcommand{\FIGj}{abs-em}
\newcommand{\TABa}{CeIII_res}

\newcommand{\TABc}{CeIII-CeIV}

\newcommand{\TABe}{exp2}
\newcommand{\TABf}{oscil_strengths}
\newcommand{\TABg}{CeIII-CeIV-offset}

\newcommand{\Luiscolor}{magenta}   

\newcommand{\FIGI}{\FIGg}
\newcommand{\FIGII}{\FIGh}
\newcommand{\FIGIII}{\FIGi}
\newcommand{\FIGIV}{\FIGa}
\newcommand{\FIGV}{\FIGj}
\newcommand{\FIGVI}{\FIGc}
\newcommand{\FIGVII}{\FIGd}
\newcommand{\FIGVIII}{\FIGf}

\newcommand{\TABI}{\TABe}
\newcommand{\TABII}{\TABa}

\usepackage[final]{graphicx}   
\usepackage{dcolumn}           
\usepackage{bm}                
\usepackage{longtable} 
\usepackage{rotating}
\usepackage{color}
\usepackage{comment}
\usepackage{epstopdf}
\usepackage[bitstream-charter]{mathdesign}
\usepackage[T1]{fontenc}

\def\abinitio{{\it ab initio}}

\def\etal{{\it et al.}}

\def\cmm1{cm$^{-1}$}

\def\nue{$\omega_{a_{1g}}$}
\def\Te{T$_{e}$}

\def\HostCsLiLuCl{Cs$_2$LiLuCl$_6$}
\def\HostCsLiYCl{Cs$_2$LiYCl$_6$}
\def\HostCsNaYCl{Cs$_2$NaYCl$_6$}
\def\HostCsNaYF{Cs$_2$NaYF$_6$}

\def\HostCsGeF{Cs$_2$GeF$_6$}

\def\SrClII{SrCl$_2$}

\def\CeIII{Ce$^{3+}$}
\def\CeIV{Ce$^{4+}$}

\def\CsI{Cs$^{+}$}

\def\LiI{Li$^{+}$}

\def\LuII{Lu$^{2+}$}
\def\LuIII{Lu$^{3+}$}

\def\NdIII{Nd$^{3+}$}

\def\PrIII{Pr$^{3+}$}

\def\UIV{U$^{4+}$}

\def\YbII{Yb$^{2+}$}

\def\Clm{Cl$^{-}$}

\def\CeClvimii{(CeCl$_6$)$^{2-}$}
\def\CeClvimiii{(CeCl$_6$)$^{3-}$}

\def\OCeiv+xiv{(OCe$_4$)$^{14+}$}
\def\OUiv+xiv{(OU$_4$)$^{14+}$}

\def\CeiiClxiimv{(Ce$_2$Cl$_{12}$)$^{5-}$}

\def\dCeCl{$d_{\rm Ce-Cl}$}
\def\dCeCle{$d_{{\rm Ce-Cl},e}$}
\def\dCeClL{$d_{{\rm Ce-Cl,L}}$}
\def\dCeClR{$d_{{\rm Ce-Cl,R}}$}
\def\dCeClAC{$d_{{\rm Ce-Cl,AC}}$}

\iftoggle{journal-like}           {\pagestyle{fancy} \lhead{\Authors}\chead{}\rhead{\myheading}\lfoot{}\cfoot{\thepage}\rfoot{} } {}
\iftoggle{journal-like-bigfigures}{\pagestyle{fancy} \lhead{\Authors}\chead{}\rhead{\myheading}\lfoot{}\cfoot{\thepage}\rfoot{} } {}

\iftoggle{usualpreprint}          { } {}
\iftoggle{journal-like}           { } {}
\iftoggle{journal-like-bigfigures}{ } {}

\begin{document}
\title{
       Intervalence Charge Transfer Luminescence: 
       The Anomalous Luminescence of Cerium-Doped  Cs$_2$LiLuCl$_6$ Elpasolite.
      }
\date{\today}

\author{Luis Seijo}
\affiliation{Departamento de Qu\'{\i}mica, 
             Universidad Aut\'onoma de Madrid, 28049 Madrid, Spain}
\affiliation{Instituto Universitario de Ciencia de Materiales Nicol\'as Cabrera,
             Universidad Aut\'onoma de Madrid, 28049 Madrid, Spain}
\author{Zoila Barandiar\'an}
\affiliation{Departamento de Qu\'{\i}mica, 
             Universidad Aut\'onoma de Madrid, 28049 Madrid, Spain}
\affiliation{Instituto Universitario de Ciencia de Materiales Nicol\'as Cabrera,
             Universidad Aut\'onoma de Madrid, 28049 Madrid, Spain}
%
\begin{abstract}

The existence of intervalence charge transfer (IVCT) luminescence is reported.
It is shown that the so called anomalous luminescence of Ce-doped elpasolite Cs$_2$LiLuCl$_6$,
which is characterized mainly by a very large Stokes shift and a very large band width,
corresponds to an IVCT emission that takes place in Ce$^{3+}$--Ce$^{4+}$ pairs,
from the $5de_g$ orbital of Ce$^{3+}$ to $4f$ orbitals of Ce$^{4+}$.
Its  Stokes shift is the sum of the large reorganization energies of the Ce$^{4+}$ and Ce$^{3+}$ centers
formed after the fixed-nuclei electron transfer and it is equal to 
 the energy of the IVCT absortion commonly found in mixed-valence compounds, which is predicted to exist in this material 
and to be slightly larger than 10000~cm$^{-1}$.
The large band width is the consequence of the large offset between the minima of the Ce$^{3+}$--Ce$^{4+}$ and
Ce$^{4+}$--Ce$^{3+}$ pairs along the electron transfer reaction coordinate.
This offset is approximately $2\sqrt{3}$ times the difference of Ce-Cl equilibrium distances
in the Ce$^{3+}$ and Ce$^{4+}$ centers.
It is shown that the energies of the peaks and the widths of IVCT absortion and emission bands
can be calculated {\em ab initio} with reasonable accuracy 
from diabatic energy surfaces of the ground and excited states and that 
these can be obtained, in turn, from independent calculations on the donor and acceptor active centers.
We obtained the energies of the Ce$^{3+}$ and Ce$^{4+}$ active centers of Ce-doped Cs$_2$LiLuCl$_6$
by means of state-of-the-art wave-function-theory spin-orbit coupling relativistic calculations 
on the donor cluster (CeCl$_6$Li$_6$Cs$_8$)$^{11+}$ and the acceptor cluster (CeCl$_6$Li$_6$Cs$_8$)$^{12+}$ 
embedded in a quantum mechanical embedding potential of the host.
The calculations provide interpretations of  unexplained experimental observations 
as due to higher energy IVCT absorptions, and allow to reinterpret others.
The existence of another IVCT emission of lower energy, at around 14000-16000~cm$^{-1}$ less than the 
$5dt_{2g}$ emission, is also predicted.

\end{abstract}
\maketitle
\section{Introduction}
\label{SEC:introduction}

Intervalence charge transfer (IVCT) is the conventional name for electron transfer between two metal sites 
differing only in oxidation state.~\cite{VERHOEVEN:96}
Therefore,
it is a particular case of metal-to-metal charge transfer (MMCT) in which the two metal ions involved in 
the redox process are identical; 
in other words, it is the homonuclear, symmetric MMCT.~\cite{IVCT-MMCT}
The basic theory for electron transfer was formulated by Marcus~\cite{MARCUS:64} and 
the systematic study of intervalence compounds~\cite{ALLEN:67,ROBIN:68} 
has played a key role in elucidating electron transfer reactions.
The electron transfer between the ground states of the two metal sites 
can be thermally induced after passing through an activated complex 
with an activation energy barrier; 
in the activated complex the electron is equally distributed among the two metallic centers.
The electron transfer can also be photoinduced. 
In this case, a fixed nuclei IVCT photon absorption takes place
that is followed by a nonradiative decay involving nuclei reorganization;
the decay passes through the activated complex and it can branch either to the original state or to the charge transfer state,
which are degenerate (Fig.~\ref{FIG:\FIGg}, red lines). 
Obviously, there cannot be any emission associated with the IVCT absorption. 
IVCT absorption explained the early observations of Werner~\cite{WERNER:96} on the dark color of substances 
containing platinum in two oxidation states
and it was found in a large number of mixed valence compounds,~\cite{ALLEN:67,ROBIN:68}
mostly involving transition metals.
The Marcus theory of electron tranfer 
was complemented by Hush~\cite{HUSH:67,HUSH:85} with a theory of heteronuclear MMCT and homonuclear IVCT
considering a two-state problem.
Piepho \etal~\cite{PIEPHO:78} formulated a vibronic model for the IVCT absorption of the two-state problem.
We are not aware of extensions of these theories to absorptions and emissions involving
higher excited states of the mixed valence compounds.

\iftoggle{journal-like}{
  \def\escalafig{0.4} 
  \begin{figure}[t] 
\resizebox{\escalafig\textwidth}{!}{
    \rotatebox{00}{\includegraphics[scale=0.6,clip]{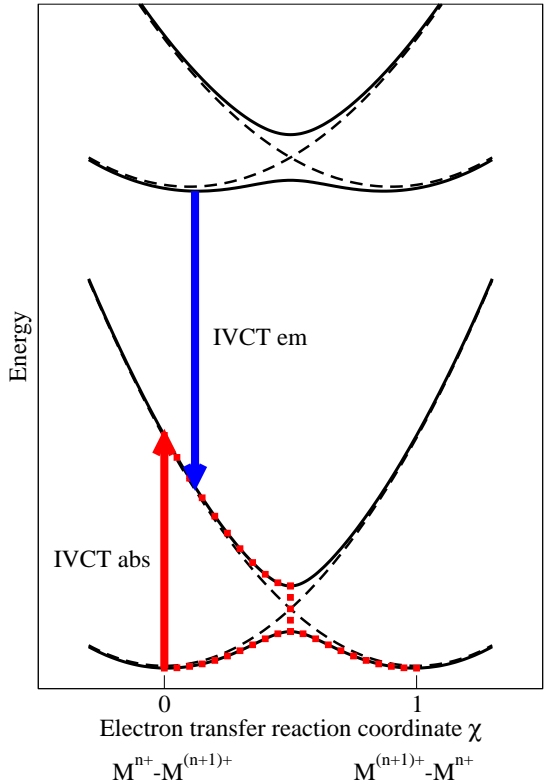}}
}
\caption{
         Schematic representation of          
         an intervalence charge transfer absorption (red arrow) 
         and emission (blue arrow),
         followed by nonradiative decay (red squares). 
         Diabatic (dashed lines) and adiabatic (full lines) energy curves
         of the ground and excited states of the 
         M$^{n+}$--M$^{(n+1)+}$ and M$^{(n+1)+}$--M$^{n+}$ pairs are shown.
}
\label{FIG:ivct}
  \end{figure}
}{}

The relevance of MMCT transitions in solids has been reviewed by Blasse,~\cite{BLASSE:91}
including IVCT absorptions in mixed-valence compounds;
most of the materials involve transition metal ions.
In f-element doped solids, MMCT between f-elements and the cations of the host
have been considered responsible for green-blue luminescence quenching and red luminescence induction 
in \PrIII-doped compounds,~\cite{BOUTINAUD:07}
and a near-IR/Visible broad absorption band in Ce-doped LaPO$_4$ 
has been ascribed to a \CeIII--\CeIV\ IVCT absorption,~\cite{VANSCHAIK:93}
for instance,
but neither heteronuclear MMCT nor homonuclear IVCT have been the subject of extensive investigations.

Although the IVCT absorption is very well known in mixed valence compounds,
IVCT luminescence transitions (Fig.~\ref{FIG:\FIGg} blue arrow) have never been reported,
to the best of our knowledge.
A series of \abinitio\ studies on lanthanide optically active centers in solids is leading us to propose that IVCT emissions
have in fact been observed in Ce-doped elpasolites and in Yb-doped fluorites,
even though they have not been identified as such.
We also expect these emissions to be present in Eu-doped solids
and in other solids doped with f-elements in which several valence states are likely to coexist. 
IVCT states are also likely to be responsible for luminescence quenching in many materials.
Here we report \abinitio\ calculations on the IVCT luminescence of Ce-doped \HostCsLiLuCl.
In a separate paper we report \abinitio\ calculations on the IVCT luminescence of Yb-doped fluorites.~\cite{BARANDIARAN:14}

Ce-doped \HostCsLiLuCl\ elpasolite is a scintillating material that presents an anomalous luminescence.~\cite{BESSIERE:06}
Such emission is excited by the $4f \rightarrow 5de_g$ absortion 
but it cannot be a usual $5de_g \rightarrow 4f$ emission because
its energy is much lower than its excitation (more than 9000~\cmm1\ lower)
and its band width is very large (full width at half maximum of 4800~\cmm1\ at room temperature).
Its intensity increases with temperature up to 250~K and decreases above it.
This anomalous luminescence was also found in Cs$_3$LuCl$_6$:Ce [Ref.~\onlinecite{DORENBOS:03:b}]
and \HostCsNaYCl:Ce [Ref.~\onlinecite{BESSIERE:04}];
in these materials it is quenched at room temperature 
but the basic mechanism is likely to be the same in all three.

In this work, 
we interpret the anomalous luminescence of \HostCsLiLuCl:Ce
as a \CeIII\ $5de_g$ $\rightarrow$ \CeIV\ $4f$ intervalence charge transfer emission 
that takes place in \CeIII--\CeIV\ active pairs.
This interpretation is based on wave function theory \abinitio\ calculations on Ce-doped \HostCsLiLuCl\
which involve:
the $4f^1$ and $5d^1$ manifolds of the \CeIII\ active center, and its $a_{1g}^1$ impurity-trapped exciton (ITE);
the \CeIV\ closed-shell center;
and the diabatic energy surfaces of the \CeIII--\CeIV\ active pair
along the electron transfer reaction coordinates of
the ground state and excited states.

The results suggest that \CeIII-\CeIV\ IVCT states might also be responsible for other not well understood features
of related materials,
like unassigned low-T excitation bands of \HostCsNaYCl:Ce [Ref.~\onlinecite{DUAN:09}]
and \HostCsNaYF:Ce [Ref.~\onlinecite{DUAN:11}],
and the anomalous behaviour of the 351~nm emission of \HostCsNaYF:Ce under various excitation wavelengths.~\cite{DUAN:11}

Also, it is shown that
the diabatic energy surfaces of the \CeIII--\CeIV\ mixed valence pairs are sufficient to
provide the basic understanding of the states involved in the relevant absorptions, emissions, and nonradiative decays.
They can be obtained from independent \abinitio\ calculations in embedded monometallic clusters 
of the oxidized and reduced species, like \CeClvimiii\ and \CeClvimii.
The adiabatic energy surfaces, 
which require much more demanding calculations on embedded dimetallic clusters like \CeiiClxiimv,
are only necessary when electronic interaction between them are
expected to be significant and accurate numerical results
are targeted.

%

\section{Potential energy surfaces of mixed valence active pairs}

The intervalence charge transfer luminescence of Ce-doped \HostCsLiLuCl\ 
takes place between states of \CeIII-\CeIV\ mixed valence active pairs.
In Secs.~\ref{SEC:adiabatic_PES} and \ref{SEC:diabatic_PES} 
we describe the basics of their adiabatic and diabatic potential energy surfaces 
(equivalent to the full and dashed lines in Fig.~\ref{FIG:\FIGg}, respectively)
and their relations.


\subsection{Adiabatic potential energy surfaces}
\label{SEC:adiabatic_PES}

Let us consider a {\em donor} $D$ and an {\em acceptor} $A$,
which in this case will be the separated embedded clusters \CeClvimiii\ and \CeClvimii.
Let us also consider a {\em donor-acceptor} pair
prior to electron transfer, $DA$, and after the electron transfer, $D^+A^-$.
In the homonuclear, symmetric MMCT case, or IVCT, $D^+$ is $A$ and $A^-$ is $D$,
so that we can call the pair after the electron transfer 
the {\em acceptor-donor} pair $AD$.
In this case, $DA$ and $AD$  will be the mixed valence embedded cluster  \CeiiClxiimv\ 
with conventional left--right distributions \CeIII--\CeIV\ and \CeIV--\CeIII\ respectively.
We are interested in the energies of the  $DA$ and $AD$ pairs  in their ground and excited states
as functions of the nuclear coordinates. 
Since the electronic spectroscopic transitions are dominated by the totally symmetric vibrational coordinates,
we will only consider the breathing modes of octahedral \CeClvimiii\ and \CeClvimii, 
so that the only vibrational degrees of freedom will be the Ce-Cl distances in the left and right components of the pairs,
$d_L$ and $d_R$.
For each nuclear configuration,
the electronic states of the pairs result from the combinations between 
the $n_D$ individual states of $D$ and the $n_A$ individual states of $A$.
In this case they will be the states of the $4f^1$, $5dt_{2g}^1$, $5de_g^1$, 
and $a_{1g,ITE}^1$ (impurity-trapped exciton) configurations of \CeClvimiii\
combined with the $A_{1g}$ closed-shell state of \CeClvimii.
Besides, since both $DA$ and $AD$ electronic configurations are possible,
the total number of adiabatic electronic states $\Phi_k$ of the mixed valence pair is $2n_Dn_A$.
Their adiabatic potential energy surfaces (equivalent to the full lines in Fig.~\ref{FIG:\FIGg}) 
will be $E_{k}(d_L,d_R)$, $k=1,2,\ldots,2n_Dn_A$.
These adiabatic energy surfaces can provide basic spectroscopic data
without requiring to solve the full vibronic problem,
like the positions of absorption and emission band maxima via the Frank-Condon approximation, 
and others.
In general, they result from direct quantum mechanical calculations on mixed valence dimers, which are highly demanding.
Depending on the number of open-shell electrons of the lanthanide, they can be extremely demanding.


\subsection{Diabatic potential energy surfaces}
\label{SEC:diabatic_PES}

\subsubsection{Definition}

Independently of the method used for the calculation of the adiabatic potential energy surfaces,
they can be considered to result from the diagonalization of a 
$(2n_Dn_A \times2n_Dn_A )$ interaction matrix.
The diagonal elements of the interaction matrix are the diabatic potential energy surfaces 
(equivalent to the dashed lines in Fig.~\ref{FIG:\FIGg})
and the off-diagonal elements are the electronic couplings.
The diabatic energy surfaces can cross each other and, contrary to the adiabatic, 
they mantain the nature of the electronic state across the crossings.
The diabatic basis is arbitrary; 
a convenient choice is the set of generalized antisymmetric product functions~\cite{MCWEENY:59,McWeeny89}
resulting from the combination of the states of $D$ and $A$.
So, from  the combination of the $i$ state of $D$, $\Phi_{Di}$, and the $j$ state of $A$, $\Phi_{Aj}$,
we will have two diabatic wavefunctions:
    one for the $ij$ state of $DA$, $M \hat A (\Phi_{Di} \Phi_{Aj})$,
and one for the $ji$ state of $AD$, $M \hat A (\Phi_{Aj} \Phi_{Di})$ 
($M$ is a normalization constant and $\hat A$ is the inter-group antisymmetrization operator~\cite{MCWEENY:59}).
Their expected values of the fixed nuclei Hamiltonian of the embedded pair $\hat{H}$, are the
two corresponding diabatic potential energy surfaces
$E_{DiAj}^{\rm diab}(d_L,d_R)$ and  $E_{AjDi}^{\rm diab}(d_L,d_R)$:
\mbox{$E_{DiAj}^{\rm diab} = \langle M\hat A (\Phi_{Di} \Phi_{Aj}) | \hat H | M\hat A (\Phi_{Di} \Phi_{Aj}) \rangle$,}
\mbox{$E_{AjDi}^{\rm diab} = \langle M\hat A (\Phi_{Aj} \Phi_{Di}) | \hat H | M\hat A (\Phi_{Aj} \Phi_{Di}) \rangle$.}
Their electronic coupling is
\mbox{$V_{DiAj,AjDi}^{\rm diab} = \langle M\hat A (\Phi_{Di} \Phi_{Aj}) | \hat H | M\hat A (\Phi_{Aj} \Phi_{Di}) \rangle$.}
Note that $E_{DiAj}^{\rm diab}(x,y) = E_{AjDi}^{\rm diab}(y,x)$.

We must remark now that,
if the full vibronic problem of the mixed valence pair needed to be solved,
both the adiabatic basis $\{ \Phi_k \}$ 
and the diabatic basis $\{ M\hat A (\Phi_{Di} \Phi_{Aj}), \, M\hat A (\Phi_{Aj} \Phi_{Di}) \}$,
would be valid alternatives 
leading to the same vibronic energies and wavefunctions.
If, instead, the focus is  on
some relevant features of the electronic spectra
like zero-phonon lines, absorption and emission band maxima and band widths, etc.,
or of the electronic states of the mixed valence pair,
like equilibrium structures, activation energy barriers, etc.,
the adiabatic energy surfaces are the valid ones and the diabatic energy surfaces are an approximation to them.
Let us briefly discuss on the limits of the diabatic approximation.

In the regions of the nuclear configuration space where the electronic couplings are not very large,
the adiabatic and diabatic energy surfaces are  close,
each adiabatic wavefunction $\Phi_k$ is very similar to one of the diabatic wavefunctions,
and a particular $D_iA_j$ or $A_jD_i$ character can be associated with it.
E.g.,
for $d_L$ and $d_R$ values respectively close 
to the Ce-Cl equilibrium distance of the \CeIII\ donor site and of the \CeIV\ acceptor site,
one of the adiabatic energies will be close to $E_{DiAj}^{\rm diab}$ and
another to $E_{AjDi}^{\rm diab}$;
$E_{DiAj}^{\rm diab}$ will be the lowest of the two 
because both the donor and the acceptor are structurally relaxed,
and $E_{AjDi}^{\rm diab}$ will be the highest of the two
because both the donor and the acceptor are structurally stressed,
each of them being in the equilibrium structure of the other.  
This is an important observation 
because the calculations of the diabatic energy surfaces,
which can be hihgly demanding, 
are always significantly less demanding than the calculations of the adiabatic ones.

In the regions of the nuclear configuration space near crossings of diabatic energy surfaces, 
e.g. near the activated complex where \mbox{$d_L \approx d_R$,}
the electronic couplings have larger effects and  produce avoided crossings.
The crossing diabatic states are then replaced by a lower adiabatic state
with a smaller thermal energy barrier, plus an upper adiabatic state  which has become
stable at the configuration where the electron is equally distributed among the two
metals (see Fig.~\ref{FIG:\FIGg}).
In these regions, the adiabatic results are necessary when quantitative energy barriers 
or quantitative nonradiative dynamics are targeted. 
Many of the spectroscopic features of the mixed valence pairs can however be addressed 
quantitatively or semiquantitatively with the diabatic energy surfaces only.
We describe next how they can be computed.

\subsubsection{Approximations}
\label{SEC:DPESapprox}

The diabatic pair energies are the sum of the donor and acceptor energies 
plus their mutual Coulomb and exchange interaction.~\cite{McWeeny89}
The latter should be almost independent of the donor and acceptor states, in general.
Hence, we can write:
\begin{eqnarray}
E_{DiAj}^{\rm diab} = E_{Di} + E_{Aj} + E^{cx}_{DiAj} \approx E_{Di} + E_{Aj} + E^{cx}_{DA}
\,.
\label{EQ:EDA}
\end{eqnarray} 
In Eq.~\ref{EQ:EDA}, $E_{Di}$ and $E_{Aj}$ include the embedding interactions of $D$ and $A$ with the crystalline environment 
of the $DA$ pair in \HostCsLiLuCl.
We aim at  computing $E_{DiAj}^{\rm diab}$ by means of embedded cluster calculations
and we can think of two alternative computational strategies:
In one of them, the symmetry reductions around $D$ and $A$ due to the presence of the other ($A$ and $D$ respectively)
are considered from the very beginning.
In the other, they are removed in a first step and they are considered later, at the same time that the electronic couplings,
i.e. when the adiabatic surfaces are calculated. 
The strength of the first approach is to be able to give energy splittings driven by symmetry lowering,
which are dependent on the distance and relative orientation between $D$ and $A$.
The strength of the second approach is to be able to give the basics of the energy surfaces 
by means of independent calculations on the embedded $D$ and $A$.
These alternatives are as follows:

{\bf 1}) One calculation {\bf 1D} of the donor \CeClvimiii\ embedded in a \HostCsLiLuCl\ lattice
in which one  \LuIII\ ion is substituted by a \CeIV\ ion 
 gives $E_{Di}^{(1)} = E_{Di} + E^{cx}_{DA}$.
Another calculation {\bf 1A} of the acceptor \CeClvimii\ embedded in a  \HostCsLiLuCl\ lattice
in which one  \LuIII\ ion is substituted by a \CeIII\ ion 
 gives $E_{1Aj}^{(1)} = E_{Aj} + E^{cx}_{DA}$.
Then, $E_{DiAj}^{\rm diab} = E_{Di}^{(1)} + E_{Aj}^{(1)} - E^{cx}_{DA}$.
Here, $E_{Di}^{(1)}$ and $E_{Aj}^{(1)}$ depend on the relative positions of $D$ and $A$.
$E^{cx}_{DA}$ is dominated by the long-range Coulomb interactions,
so that, except for short donor-acceptor distances, 
$E^{cx}_{DA} \approx - (q_D \times q_A) e^2/d_{DA}$
$= - (3 \times 4) e^2/d_{DA}$.
According to this alternative we have:
\begin{eqnarray}
E_{DiAj}^{\rm diab}(d_L,d_R) = E_{Di}^{(1)}(d_L) + E_{Aj}^{(1)}(d_R) + \Delta E^{cx}_1(d_{DA}) 
\,,
\nonumber\\
E_{AjDi}^{\rm diab}(d_L,d_R) = E_{Ai}^{(1)}(d_L) + E_{Dj}^{(1)}(d_R) + \Delta E^{cx}_1(d_{DA}) 
\,,
\end{eqnarray} 
with 
\begin{eqnarray}
\Delta E^{cx}_1(d_{DA}) = - E^{cx}_{DA}(d_{DA}) \approx - (q_D \times q_A) e^2/d_{DA}
\,.
\label{EQ:E01}
\end{eqnarray} 

{\bf 2}) One calculation {\bf 2D} of the donor \CeClvimiii\ embedded in a \HostCsLiLuCl\ lattice
 gives $E_{Di}^{(2)} = E_{Di} + E^{cx}_{DC}$,
where $E^{cx}_{DC}$ stands for the Coulomb and exchange interaction energy between the donor \CeClvimiii\ 
and the cluster with the original host cation C (\LuIII\ in this case), (LuCl$_6)^{3-}$.
Another calculation {\bf 2A} of the acceptor \CeClvimii\ embedded in a  \HostCsLiLuCl\ lattice
 gives $E_{Aj}^{(2)} = E_{Aj} + E^{cx}_{AC}$,
where $E^{cx}_{AC}$ stands for the Coulomb and exchange interaction energy between the acceptor \CeClvimii\ 
and the cluster with the original host cation, (LuCl$_6)^{3-}$.
Then, $E_{DiAj}^{\rm diab} = E_{Di}^{(2)} + E_{Aj}^{(2)} - E^{cx}_{DC} - E^{cx}_{AC} + E^{cx}_{DA}$.
Here, $E_{Di}^{(2)}$ and $E_{Aj}^{(2)}$ are independent of the relative positions of $D$ and $A$.
Except for short cation-cation distances,
$- E^{cx}_{DC} - E^{cx}_{AC} + E^{cx}_{DA}$ 
$\approx (- q_D \times q_C - q_A \times q_C + q_D \times q_A) e^2/d_{DA}$
$= (- 3 \times 3 - 4 \times 3 + 3 \times 4) e^2/d_{DA}$
$=- (3\times3) e^2/d_{DA}$.
According to this alternative we have:
\begin{eqnarray}
E_{DiAj}^{\rm diab}(d_L,d_R) = E_{Di}^{(2)}(d_L) + E_{Aj}^{(2)}(d_R) + \Delta E^{cx}_2(d_{DA})
\,,
\nonumber\\
E_{AjDi}^{\rm diab}(d_L,d_R) = E_{Aj}^{(2)}(d_L) + E_{Di}^{(2)}(d_R) + \Delta E^{cx}_2(d_{DA})
\,,
\end{eqnarray} 
with 
\begin{eqnarray}
\Delta E^{cx}_2(d_{DA}) &=& E^{cx}_{DA}(d_{DA}) - E^{cx}_{DC}(d_{DA}) - E^{cx}_{AC}(d_{DA})
\nonumber\\
&\approx& (q_D \times q_A - (q_D+q_A) \times q_C) e^2/d_{DA}
\,.
\label{EQ:E02}
\end{eqnarray} 

Summarizing,
the diabatic potential energy surfaces will be given by:
\begin{eqnarray}
E_{DiAj}^{\rm diab}(d_L,d_R) =  E_{Di}(d_L) + E_{Aj}(d_R) + E_0(d_{DA})
\,,
\nonumber\\
E_{AjDi}^{\rm diab}(d_L,d_R) =  E_{Aj}(d_L) + E_{Di}(d_R) + E_0(d_{DA})
\,,
\label{EQ:DiabaticPES}
\end{eqnarray} 
with donor and acceptor energies $E_{Di}$ and $E_{Aj}$ obtained 
 in embedded cluster calculations {\bf 1D} and {\bf 1A}
and the term $E_0$, which is common to the $AD$ and $DA$ energy surfaces, given by Eq.~\ref{EQ:E01} (alternative {\bf 1}),
or with donor and acceptor energies obtained 
 in embedded cluster calculations {\bf 2D} and {\bf 2A}
and the common term $E_0$ given by Eq.~\ref{EQ:E02} (alternative {\bf 2}).
Alternative {\bf 1} has into account the effects of charge substitutions in the original lattice
on the energy levels of $D$ and $A$. 
In the present case, the most important ones are expected to be the splittings produced on the  \CeClvimiii\ levels.
They should be responsible for fine features of the spectra, but not for the number and positions of the main
absorption and emission bands.
In any case, this alternative implies site symmetry reduction, which may add significant computatioanl effort.
Alternative {\bf 2} neglects these effects.
It has, however, an important computational advantage:
the diabatic potential energy surfaces of the mixed valence pairs are computed using the energy curves 
of the donor and acceptor centers embedded in the original host lattice,
i.e. of the clusters  \CeClvimiii\ and \CeClvimii\ embedded in \HostCsLiLuCl\ in our case.
%
We must remark that, regardless of the alternative used, 
the term $E_0(d_{DA})$ is common to the $DA$ and $AD$ energy surfaces and to all states of both.
Its effect is a common shift of all of them and, consequently, it does not contribute to energy differences between them.
%
In this work we have adopted alternative 2.

\subsubsection{Topology}

\iftoggle{journal-like}{
  \def\escalafig{0.45} 
  \begin{figure}[b] 
\resizebox{\escalafig\textwidth}{!}{
    \rotatebox{00}{\includegraphics[scale=1,clip]{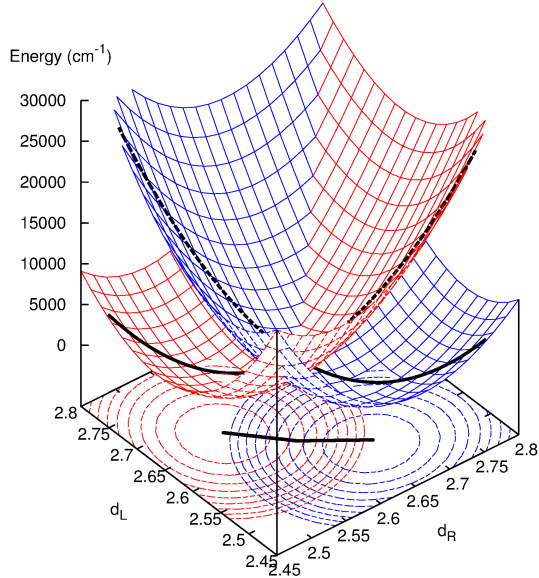}}
}
\caption{
         Diabatic potential energy surfaces
         of the $DA$ mixed valence pair \CeClvimiii-\CeClvimii\ (red) 
         and the $AD$ mixed valence pair \CeClvimii-\CeClvimiii\ (blue) 
         in their ground states.
         $d_L$ and $d_R$ stand for the Ce-Cl distances in the left and right clusters, respectively.
         Black straight lines on the $d_L$-$d_R$ plane connect 
         the position of activated complex of the electron transfer reaction 
         with the equilibrium positions of the pairs; 
         these straight lines define the electron transfer reaction coordinate.
         The lowest (full lines) and highest (dashed lines) diabatic energies along the reaction coordinate
         are shown on the surfaces.
}
\label{FIG:gs_surfaces}
  \end{figure}
}{}

The diabatic potential energy surfaces of the $DA$ and $AD$ mixed valence pairs in their ground states,
$E_{D0A0}^{\rm diab}(d_L,d_R)$ and $E_{A0D0}^{\rm diab}(d_L,d_R)$,
are shown in Fig.~\ref{FIG:\FIGh}  for our case. 
An equivalent result is found for any  pair of $E_{DiAj}^{\rm diab}(d_L,d_R)$ and $E_{AjDi}^{\rm diab}(d_L,d_R)$ 
excited state surfaces,
so that the following discussion also holds for them.
The diabatic energy surfaces correspond to the \CeClvimiii--\CeClvimii\ pair (red surface)
and to the \CeClvimii--\CeClvimiii\ pair (blue surface), respectively.
Their respective minima are found where the left and right distances $d_L$ and $d_R$ take the values of the
donor \CeClvimiii\ and acceptor \CeClvimii\  ground state Ce-Cl equilibrium distances, 
$d_{eD}$ and $d_{eA}$ (2.660~\AA\ and 2.542~\AA\ in our case; see below), 
and viceversa:  ($d_{eD},d_{eA}$) and ($d_{eA},d_{eD}$).
The energies at the minima are equal:
$E_{D0A0}^{\rm diab}(d_{eD},d_{eA}) = E_{A0D0}^{\rm diab}(d_{eA},d_{eD}) = E_{e,D0A0}$.

The $DA$ and $AD$ diabatic surfaces are degenerate along the $d_L$=$d_R$ line.
So, 
the diabatic activated complex of the $D + A \rightarrow A + D$ electron transfer reaction,
which is the crossing point between the two surfaces with lowest energy,
can be found by minimization of any of the two surfaces along the $d_L$=$d_R$ line.
(Strictly speaking, the activated complex is the full $d_L$=$d_R$ line,
but here we will use the term only for its most stable structure.) 
In the diabatic activated complex, 
which is expected to be close to the adiabatic,
the left and right distances take a common value $d_{ac}$.
In our case, the minimum is found 
at a common Ce-Cl distance of the donor \CeClvimiii\ and the acceptor \CeClvimii\  
of $d_{ac}$=2.599~\AA.

A fairly good estimation of $d_{ac}$ can be obtained from the harmonic approximation of the diabatic surfaces.
In this approximation, the activated complex is found at
$d_{ac} = (\omega_D^2 d_{eD} + \omega_A^2 d_{eA})/(\omega_D^2 + \omega_A^2)$,
$\omega_D$ and $\omega_A$ being the vibrational frequencies of $D$ and $A$ ground states respectively.
In other words,
$d_{ac}$ would be the $\omega^2$-weighted average of donor and acceptor equilibrium distances.
If equal values for the vibrational frequencies of donor and acceptor are assumed,
as it is done in the vibronic model of Piepho \etal,~\cite{PIEPHO:78} 
$d_{ac}$ is the simple average, $d_{ac} = (d_{eD} + d_{eA})/2$.
Since the acceptor has a higher force constant than the donor,
$d_{ac}$ should be close to the average of $d_{eD}$ and $d_{eA}$ but closer to $d_{eA}$ than to $d_{eD}$.
The values $\omega_D$=313~\cmm1\ and $\omega_A$=355~\cmm1\ found in the present \abinitio\ calculations (see below) 
 give $d_{ac}$=2.594~\AA\ for the different-frequencies harmonic approximation
and  $d_{ac}$=2.601~\AA\ for the equal-frequencies harmonic approximation.
Both of them are close to the found $d_{ac}$=2.599~\AA, although, interstingly, the latter is closer.
It is a manifestantion of the fact that anharmonicity increases the Ce-Cl distance of the activated complex,
as it does with the equilibrium distances of donor and acceptor.
It means that there is a compensation of the errors due to assuming equal force constants and neglecting anharmonicity.

The diabatic electron transfer activation energy is
$E_{D0A0}^{\ddagger\,\rm diab} = E_{D0A0}^{\rm diab}(d_{ac},d_{ac}) - E_{e,D0A0}$.
It is independent of the $d_{DA}$ distance between donor and acceptor
(within approximation 2 of Sec.~\ref{SEC:DPESapprox})
and it is an upper bound of the adiabatic activation energy,
which is $d_{DA}$-dependent.

The ground state diabatic electron transfer reaction coordinate $Q_{et}$ can be aproximated with the straight lines that connect the
activated complex $(d_{ac},d_{ac})$ with the two minima $(d_{eD},d_{eA})$ and $(d_{eA},d_{eD})$.
This reaction coordinate is represented in Fig.~\ref{FIG:\FIGh} in the $d_L$-$d_R$ plane.
The $DA$ pair has the lowest diabatic energy in the left side of the activated complex ($d_L > d_R$)
and the $AD$ pair in the right side ($d_L < d_R$),
in correspondance with the larger size of $D$ at equilibrium.
The lowest and highest diabatic energies along the reaction coordinate are shown in Fig.~\ref{FIG:\FIGh}
with full and dashed lines, respectively, drawn on the surfaces.
Since these lines contain the most interesting information of the diabatic energy surfaces,
it is convenient to plot them in energy diagrams along the reaction coordinate
(as the dashed lines in Fig.~\ref{FIG:\FIGg})
instead of the more cumbersome energy surfaces,
i.e. $E$ vs. $Q_{et}$ instead of $E$ vs. ($d_L,d_R$).

For a precise definition of $Q_{et}$, we can recall that
the changes of the Ce-Cl distances in the left and right clusters $d_L$ and $d_R$ 
along the reaction coordinate fulfil
\begin{eqnarray}
  d_L-d_{ac} = m(d_R-d_{ac}) \;\left\{ 
    \begin{array}{cl} 
        m=\frac{d_{eD}-d_{ac}}{d_{eA}-d_{ac}} & : d_L \ge d_R \\ 
        m=\frac{d_{eA}-d_{ac}}{d_{eD}-d_{ac}} & : d_L \le d_R
    \end{array} \right.
\,.
\end{eqnarray} 
Then, the normal reaction coordinate can be written as
\begin{eqnarray}
  Q_{et} = \frac{1}{\sqrt{1+m^2}} \left( Q_R + m Q_L \right)
\,,
\end{eqnarray} 
$Q_L$ and $Q_R$ being the normal breathing modes of the left and right CeCl$_6$ moieties
with respect to their structures in the activated complex:
\begin{eqnarray}
  Q_L = \frac{1}{\sqrt{6}} \left( \delta_{Cl_{L1}} + \delta_{Cl_{L2}} + \ldots +  \delta_{Cl_{L6}} \right) 
\,,
\nonumber\\
  Q_R = \frac{1}{\sqrt{6}} \left( \delta_{Cl_{R1}} + \delta_{Cl_{R2}} + \ldots +  \delta_{Cl_{R6}} \right) 
\,,
\end{eqnarray}
which have been expressed in terms of the displacements 
$\delta_{Cl_{Lk}}$ and $\delta_{Cl_{Rk}}$
of the chlorine atoms in the left and right CeCl$_6$ moieties
away from their respective Ce atoms,
starting from the positions they occupy in the activated complex.
A graphical representation of the Cl displacements along $Q_{et}$ is shown in Fig.~\ref{FIG:\FIGi}.
\iftoggle{journal-like}{
  \def\escalafig{0.55} 
  \begin{figure}[h!] 
\resizebox{\escalafig\textwidth}{!}{
  \begin{tabular}{c}
    \rotatebox{00}{\includegraphics[scale=1,clip]{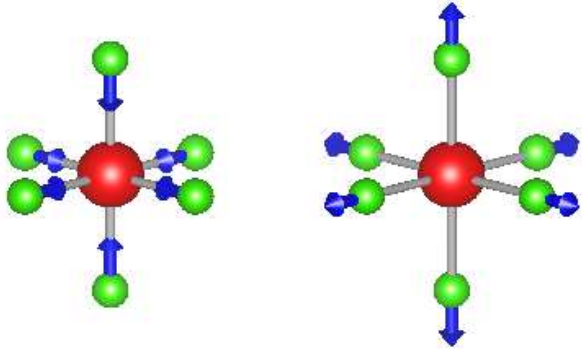}}
\\[-1cm]\mbox{}
  \end{tabular}
}
\caption{
         Displacements of the Cl atoms of the CeCl$_6$-CeCl$_6$ moieties
         along the electron transfer reaction coordinate $Q_{et}$.
}
\label{FIG:Qet}
  \end{figure}
}{}
Since the left and right cluster breathings imply
$\delta_{Cl_{L1}} = \delta_{Cl_{L2}} = \ldots = d_L - d_{ac}$ and
$\delta_{Cl_{R1}} = \delta_{Cl_{R2}} = \ldots = d_R - d_{ac}$,
we can write 
\begin{eqnarray}
  Q_L = \sqrt{6} (d_L - d_{ac})
\,,
\nonumber\\
  Q_R = \sqrt{6} (d_R - d_{ac})
\,,
\end{eqnarray} 
and
\begin{eqnarray}
  Q_{et} &=& 
             \sqrt{{6}/{(1+m^2)}} \left[ (d_R - d_{ac}) + m (d_L - d_{ac}) \right]
\nonumber\\
         &=& \sqrt{6(1+m^2)} (d_R - d_{ac})
\nonumber\\
         &=& \sqrt{6(1+m^2)} (d_L - d_{ac})/m
\,.
\label{EQ:Qet}
\end{eqnarray} 

We may mention the relationship betweem this reaction coordinate and the one of 
the vibronic model of Piepho \etal~\cite{PIEPHO:78}
The latter correspons to $d_{ac} = (d_{eD} + d_{eA})/2$, which implies $m=-1$ and gives
 $Q_{et} = (Q_R-Q_L)/\sqrt{2} $ = \mbox{$ \sqrt{3}(d_R-d_L)$}. 
We observed that this is a rather good approximation in our case.

\section{Details of the quantum mechanical calculations}
\label{SEC:Details}
\def\Donor{(CeCl$_6$Li$_6$Cs$_8$)$^{11+}$}
\def\Acceptor{(CeCl$_6$Li$_6$Cs$_8$)$^{12+}$}
\def\Host{\HostCsLiLuCl}

In this Section we describe the details of the quantum mechanical calculations of the 
$E_{Di}$(\dCeCl)  and $E_{Aj}$(\dCeCl) components of the mixed valence pair energies in 
Eq.~\ref{EQ:DiabaticPES}.
As donor $D$ and acceptor $A$ we adopted, respectively, the \Donor\ and \Acceptor\ clusters.
We performed \abinitio\ wave function theory embedded cluster quantum chemical calculations
on these  clusters embedded in the \Host\ elpasolite host
with the MOLCAS suite of programs.~\cite{MOLCAS}
The calculations include bonding interactions, static and dynamic electron correlation effects, 
and scalar and spin-orbit coupling relativistic effects within the clusters,
which are treated at a high theoretical level.
They also include Coulomb, exchange, and Pauli repulsion interations between the host and the clusters,
which are considered at a lower theoretical level by means of a quantum mechanical embedding potential.
Electron correlation effects between the cluster and the host are excluded from these calculations.

The embedded cluster calculations are two-step spin-orbit coupling 
SA-CASSCF/MS-CASPT2/RASSI-SO DKH calculations.
In the first step, we used the many-electron  
scalar relativistic second-order Douglas-Kroll-Hess (DKH) Hamiltonian.~\cite{DOUGLAS:74,HESS:86}
In \Host:\Donor,
we performed
state-average complete-active-space self-consistent-field~\cite{ROOS:80,SIEGBAHN:80,SIEGBAHN:81} (SA-CASSCF) calculations
with the active space that results from distributing one active electrons in 13 active molecular orbitals
with main character Ce $4f$, $5d$, and $6s$.
The equivalent calculation in \Host:\Acceptor\ is a closed-shell Hartree-Fock SCF calculation.
These calculations provided occupied and empty molecular orbitals to feed subsequent 
multi-state second-order perturbation theory calculations (MS-CASPT2),~\cite{ANDERSSON:90,ANDERSSON:92,ZAITSEVSKII:95,FINLEY:98} 
where the dynamic correlation of the Ce $5s,5p$, Cl $3s,3p$, and Cs $5p$ electrons
was taken into account.
We used the standard IPEA value (0.25~au).~\cite{GHIGO:04}

In the second step, 
we included spin-orbit coupling effects by adding the
Atomic Mean Field Integrals approximation (AMFI) 
of the DKH spin-orbit coupling operator~\cite{HESS:96} to the 
scalar relativistic Hamiltonian.
In this step, we used the spin-free-state-shifting operator as a means to combine
spin-orbit couplings calculated with statically correlated wave functions and
spin-orbit free energies calculated with dynamic correlation~\cite{LLUSAR:96}
and, accordingly, we performed
restricted-active-space state-interaction spin-orbit calculations 
(RASSI-SO)~\cite{MALMQVIST:02,PAULOVIC:03}
with the transformed CASSCF wave functions (first-order wave functions of the MS-CASPT2 method)  
and the MS-CASPT2 energies.

We considered $O_h$ nuclear configurations of the clusters with varying Ce-Cl distance.
The calculations were performed using the abelian $D_{2h}$ symmetry group. 
The spin-orbit coupling states of the donor \Donor\ are the following Kramers doublets:
For the {\em ungerade} $4f^1$ configuration,
one double degenerate    $\Gamma_{6u}$ of $^2T_{1u}$ character,
two double degenerate    $\Gamma_{7u}$ of $^2A_{2u}$ and $^2T_{2u}$ characters, and
two quadruple degenerate $\Gamma_{8u}$ of $^2T_{1u}$ and $^2T_{2u}$ characters;
for the {\em gerade} $5d^1$ and $6s^1$ configurations,
one double degenerate    $\Gamma_{6g}$ of $^2A_{1g}$ character,
one double degenerate    $\Gamma_{7g}$ of $^2T_{2g}$ characters, and
two quadruple degenerate $\Gamma_{8g}$ of $^2T_{2g}$ and $^2E_{g}$ characters.
The closed-shell acceptor \Acceptor\ has one $A_{1g}$ state of $^1A_{1g}$ character.

We used the following gaussian basis set:
All-electron atomic natural orbital (ANO) relativistic basis sets for 
cerium,~\cite{ROOS:08} chlorine,~\cite{ROOS:05} and lithium,~\cite{ROOS:03}
with respective contractions
($25s22p15d11f4g2h$)/[$9s8p5d4f3g2h$],
($17s12p5d4f$)/[$5s4p2d1f$], and
($14s9p$)/[$4s3p$].
For cesium, we used a
[Cd]-core \abinitio\ model potential (AIMP)~\cite{HUZINAGA:87}
that was obtained in this work from the \mbox{$1s-5s$},\mbox{$2p-4p$},\mbox{$3d-4d$} 
relativistic core ANOs of Ref.~\onlinecite{ROOS:03},
and a 
valence basis set from the same reference with the contraction 
($26s22p15d$)/[$6s5p2d$].

The Hamiltonian of the \Donor\ and \Acceptor\   clusters was supplemented
with the AIMP embedding potential~\cite{BARANDIARAN:88}
of \Host, 
which has been obtained in this work.
This embedding potential is made  of:  
1) total-ion embedding AIMPs representing \CsI, \LiI, and \LuIII\ cations and \Clm\ anions,
which are located at experimental sites of the \Host\ lattice,
within a cube made of $2 \times 2 \times 2$ unit cells and centered on \LuIII,
and 
2) a set of 25271 additional point charges situated at lattice sites,
generated by the zero-multipole method of Gell\'e and Lepetit,~\cite{GELLE:08}
which closely reproduce the Ewald potential~\cite{EWALD:21} 
within the clusters.
The experimental crystal structure of \HostCsLiLuCl\ (Ref.~\onlinecite{MEYER:86})
is the following:
Space group number 225, Fm3m cubic; 
lattice constant $a$=10.409~\AA, x(Cl)=0.2483.
The embedding AIMPs have been obtained in 
self-consistent embedded-ions (SCEI)~\cite{SEIJO:91} Hartree-Fock (HF) calculations.

The [Cd]-core AIMP of Cs and the embedding AIMP of \Host\ produced in this work are 
available from the authors~\cite{AIMP-EMBEDDING-DATA}
and are included in the supplementary material.

\section{Results and discussion}
\label{SEC:results}
\def\temission{$5dt_{2g} \rightarrow 4f$}
\def\eemission{$5de_{g} \rightarrow 4f$}

In this Section we discuss the results of the calculations, 
which we use for a better understanding of the excited states of Ce-doped \HostCsLiLuCl.
We ultimately show that the never reported IVCT luminescence has in fact been observed in this material. 
For convenience,
we summarize first the experimental energies of the excited states in Sec.~\ref{SEC:Experiments}.
We discuss in Sec.~\ref{SEC:CeIII} the states that are associated
with
the \CeIII\ active center and
in Sec.~\ref{SEC:Pair} those associated 
with
the \CeIII-\CeIV\ active pairs,
which are ultimately responsible for the IVCT luminescence.

\subsection{Excited states of Ce-doped \HostCsLiLuCl}
\label{SEC:Experiments}
The experimental energies of the excited states of Ce-doped \HostCsLiLuCl\ 
after Ref.~\onlinecite{BESSIERE:06}
are summarized in the first four columns of Table~\ref{TAB:\TABI}.
In the last three columns we summarize the results of this work,
which are discussed below in detail.
\iftoggle{journal-like}{
  \begin{widetext}\begin{center} 
  \begin{table}[h]
\caption{
Excited states of Ce-doped \HostCsLiLuCl\
after the experiments from Reference~\onlinecite{BESSIERE:06}
and this work.
}
\label{TAB:exp2}
\begin{ruledtabular}
\begin{tabular}{ccc  c ccc}
\multicolumn{4}{c}{Experiment} & \multicolumn{3}{c}{This work}   \\
Feature & nm & \cmm1\  & Assignment & Assignment & Process & \cmm1\    \\ \hline
\\
\multicolumn{7}{l}{Emission spectrum peaks (with $4f \rightarrow 5de_g$ excitation) }\\
E1  & 405,372     & 24700,26900 & $5dt_{2g} \rightarrow 4f$ $(^2F_{7/2},^2F_{5/2})$
    & same    & 2 & 20200-23300 \\  
E2  & 275         & 36400       & anomalous state $\rightarrow 4f$
    & IVCT \CeIII $5de_{g}$~$\rightarrow$~\CeIV$_{\rm sts}$ $4f$           & 7 & 32000-35300 \\  
    & (257-293)\footnotemark[1] & (34100-38900)\footnotemark[1] &  \\
\\
\multicolumn{7}{l}{Excitation spectrum peaks (monitoring $5dt_{2g} \rightarrow 4f$ emission)}\\
A1  & >330        & <30300      & $4f \rightarrow 5dt_{2g}$
    & same    & 1 & 24500, 25700 \\
A2~\footnotemark[2] & 216.5,210 & 46190,47600 & $4f \rightarrow 5de_{g}$ (JT split) 
    & same    & 3 & 46100    \\
                                                                     &&&&&& (45300,46900)~\tablenotemark[3]    \\
A3  & 303         & 33000         & unassigned        
    & IVCT \CeIII $4f$~$\rightarrow$~\CeIV$_{\rm sts}$ $5dt_{2g}$          & 8 & 31010,32170 \\
A4  & 280         & 35700         & lower symmetry \CeIII\         \\
A5  & 264,243     & 37900,41200   & lower symmetry \CeIII\        \\
A6  & 198         & 50500         & host self-trapped exciton \\      
A7  & 182         & 54900         & impurity-trapped exciton   
    & IVCT \CeIII $4f$~$\rightarrow$~\CeIV$_{\rm sts}$ $5de_{g}$           & 6 & 56240 \\       
A8  & 176         & 56800         & free excitons                
    & $4f \rightarrow a_{1g,ITE}$                                          & 5 & 59190 \\
\\
\multicolumn{7}{l}{Excitation spectrum peaks (monitoring the anomalous emission) }\\
A2  & 216.5,210   & 46190,47600  & $4f \rightarrow 5de_{g}$ (JT split) 
    & same    & 3 & 46100    \\
                                                                     &&&&&& (45300,46900)~\tablenotemark[3]    \\
A9  & 184         & 54300        &    
    & IVCT \CeIII $4f$~$\rightarrow$~\CeIV$_{\rm sts}$ $5de_{g}$           & 6 & 56240 \\       
A10 & 174         & 57500        &               
    & $4f \rightarrow a_{1g,ITE}$                                          & 5 & 59190 \\
\\
\multicolumn{7}{l}{Non-observed transitions }\\
 & & & & $5de_{g} \rightarrow 4f$ emission    & 4 & 43000-46030 \\ 
 & & & & IVCT emission \CeIII $5dt_{2g}$~$\rightarrow$~\CeIV$_{\rm sts}$ $4f$ &   & 13500-16800 \\  
 & & & & IVCT absorption \CeIII $4f$~$\rightarrow$~\CeIV$_{\rm sts}$ $4f$ & 9 & 10000-13300 \\    
\end{tabular}
\end{ruledtabular}
\footnotetext[1]{Full width at half maximum at 290~K.}
\footnotetext[2]{Absent at low temperature.}
\footnotetext[3]{A 1600~\cmm1\ splitting between the two peaks of the Jahn-Teller split absorption
is calculated at room temperature out of $\Delta E_{\rm peaks}=2\sqrt{2E_{JT}kT}$ (Reference~\onlinecite{BERSUKER:84}).}

  \end{table}
  \end{center}\end{widetext}
}{}
The manifold of excited states is made of:
states associated with the $4f^1$, $5dt_{2g}^1$, and  $5de_{g}^1$ configurations of \CeIII,
an impurity-trapped exciton (ITE) also associated with \CeIII,
and self-trapped exciton and free exciton host states.
Besides them, 
an anomalous state has been found neccesary to explain an anomalous emission.~\cite{BESSIERE:06}
Interesting features of the anomalous emission are:
it is excited by $4f \rightarrow 5de_g$ absortions but not by excitations to the host conduction band,
its energy is much lower than its excitation (large Stokes shift)
and higher than the $4f \rightarrow 5dt_{2g}$ absortions,
it is a much wider band than the regular \CeIII\ $5d \rightarrow 4f$ emissions,
and its intensity increases with temperature up to 250~K and decreases above this.
In order to understand these properties, 
the existence of an anomalous state that emits to the $4f$ ground state was suggested.~\cite{BESSIERE:06}
This state would have an energy itermediate between the $5dt_{2g}$ and  $5de_{g}$ manifolds
and it would be populated after an auto-ionization of the electron from $5de_{g}$ to the conduction band
such that it would remain near Ce.
The true nature of the anomalous state ramained not fully clear.~\cite{BESSIERE:06}

\subsection{\CeIII\ active center}
\label{SEC:CeIII}
The energy levels of the \Donor\ cluster embedded in \Host\
resulting from the spin-orbit coupling \abinitio\ calculation
are represented in Fig.~\ref{FIG:\FIGIV} as a function of the Ce-Cl distance \dCeCl\ along the $a_{1g}$ breathing mode
of the CeCl$_6$ moiety.
\iftoggle{journal-like}{
  \def\escalafig{0.4} 
  \begin{figure}[b] 
\resizebox{\escalafig\textwidth}{!}{
  \begin{tabular}{c}
    \rotatebox{00}{\includegraphics[scale=1,clip]{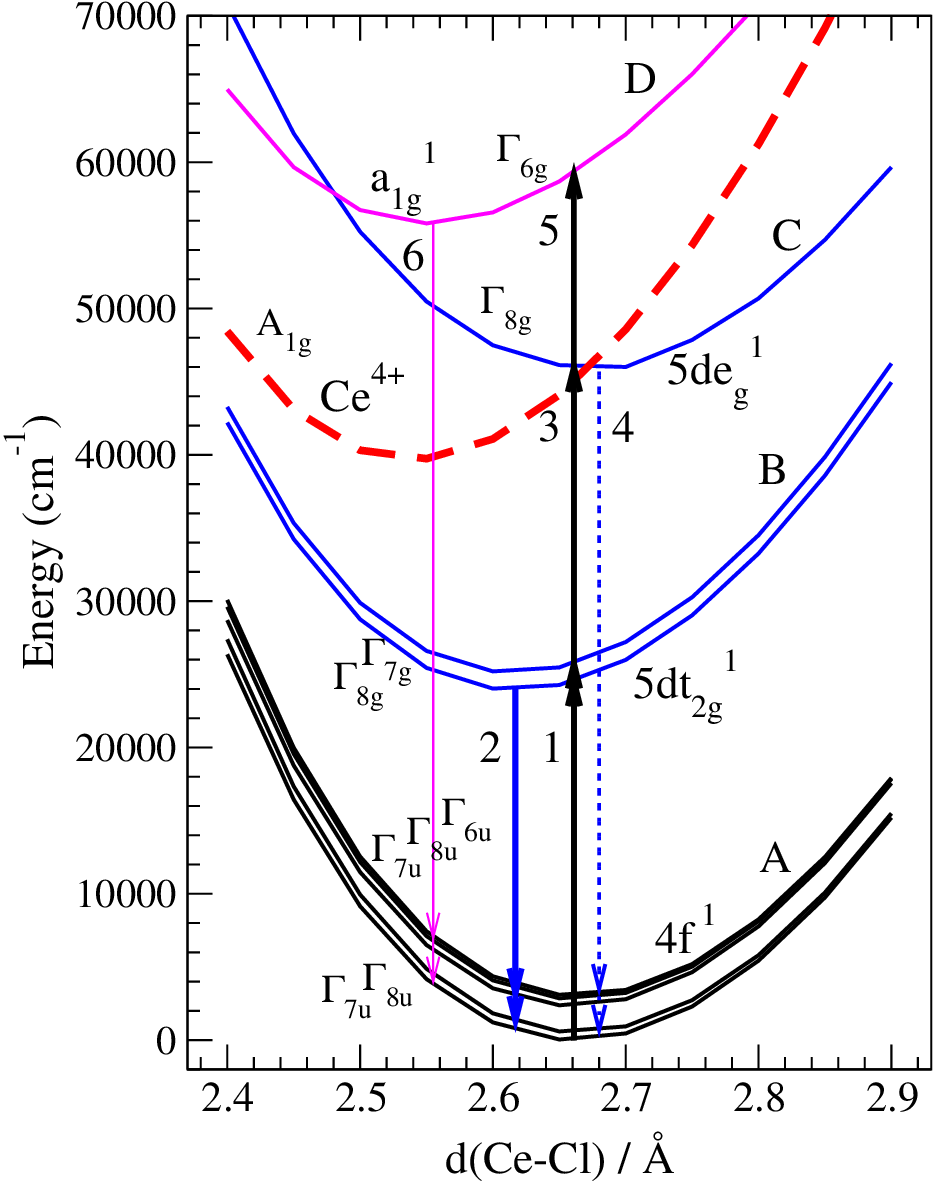}}
  \end{tabular}
}
\caption{
         Energy diagram of the \CeIII\ active center:
         Total energies of the stated of the \Donor\ cluster embedded in \HostCsLiLuCl\
         vs. the Ce-Cl distance along the totally symmetric breathing vibrational mode.
         Spin-orbit coupling RASSI calculations. 
         Energy scale relative to the ground state energy at equilibrium.
         The energy of the ground state of the \CeIV\ center in the same scale,
         as corresponds to the embedded \Acceptor\ cluster,
         is also shown (dashed red line).
         See text.
}
\label{FIG:CeIII}
  \end{figure}
}{}
We include in the Figure the total energy of the \Acceptor\ embedded cluster
in the same energy scale. 
This means that the energy difference between the $A_{1g}(^1A_{1g})$ ground state of \CeIV\ center (dashed red line)
and the $1\Gamma_{7u}$($^2A_{2u}$) state of \CeIII\ center (lowest full black line)
corresponds to the ionization potential of \CeIII, not to the conduction band of the host, but to the vacuum.
Equilibrium Ce-Cl bond lengths and breathing mode vibrational frequencies of all states,
 adiabatic transition energies (minimum-to-minimum),
and vertical (or Frank-Condon) transitions energies calculated at significant values of \dCeCl,
are given in Table~\ref{TAB:\TABII}.

The calculations give standard $4f^1$ and $5d^1$ manifolds, 
like  in \HostCsNaYCl:\CeIII\ (Ref.~\onlinecite{TANNER:03}),
and a high energy impurity-trapped exciton state,
like those previously found in \HostCsGeF:\UIV\ (Ref.~\onlinecite{ORDEJON:07})
and \SrClII:\YbII\ (Ref.~\onlinecite{SANCHEZ-SANZ:10:a}).
They support the assignments of the so called normal features of the excitation and emission spectra made
in Ref.~\onlinecite{BESSIERE:06}.

The \mbox{$4f \rightarrow 5dt_{2g}$} absorption
and \mbox{$5dt_{2g} \rightarrow 4f$} emission 
({\bf 1} and {\bf 2} in Fig.~\ref{FIG:\FIGIV} and Table~\ref{TAB:\TABII})
are calculated at 24500-25700~\cmm1\ and 20200-23350~\cmm1\ respectively.
The emission is split in two bands, 
one with three components peaking at 20200-21000~\cmm1\ due to the crystal-field splitting of $4f^1$-$^2F_{7/2}$,
and anoher with two components peaking at 22750-23350~\cmm1\ due to the crystal-field splitting of $4f^1$-$^2F_{5/2}$.
In Fig.~\ref{FIG:\FIGj}a we show  
the simulation of the shape of these absortion (blue lines) and emission (green lines) bands.
They have been calculated using
the semiclassical time-dependent approach of Heller~\cite{HELLER:75,HELLER:81,ZINK:91:b}
assuming harmonic vibrations on the initial and final electronic states.
We used a common vibrational frequency of 316~\cmm1\
and the oscillator strengths computed at the respective equilibrium geometries of the ground state (absorption)
and the respective excited states (emissions), which are shown in Table~\ref{TAB:\TABf} of Additional Material.
Experimentally,
\mbox{$4f \rightarrow 5dt_{2g}$} was found below 30300~\cmm1\ in the excitation spectrum
(its maximum was not reported)
and \mbox{$5dt_{2g} \rightarrow 4f$} split in two bands peaking at 24700 and 26900~\cmm1\ in the emission spectrum 
(A1 and E1 in Table~\ref{TAB:\TABI}).

The \mbox{$4f \rightarrow 5de_g$} absorption
({\bf 3} in Fig.~\ref{FIG:\FIGIV} and Table~\ref{TAB:\TABII})
is calculated as centered at 46100~\cmm1,
split by an $E \otimes e_g$ Jahn-Teller coupling with Jahn-Teller energy $E_{JT}$=1670~\cmm1.
In the excitation spectrum,
this transition was found as a Jahn-Teller split band with peaks at 46190 and 47600~\cmm1\
(A2 in Table~\ref{TAB:\TABI}).
The position of the band is well reproduced and its splitting corresponds to a Jahn-Teller energy not far from the calculated one.
The \mbox{$5de_g \rightarrow 4f$} emission  
({\bf 4} in Fig.~\ref{FIG:\FIGIV} and Table~\ref{TAB:\TABII})
is calculated split in two bands: one with peaks at 43000-43700~\cmm1\ and another with peaks at 45500-46000~\cmm1.
In Fig.~\ref{FIG:\FIGj}b we show  
the simulation of the shape of these absortion (blue lines) and emission (green lines) bands,
without considering the effects of the Jahn-Teller coupling 
and using the oscillator strengths of Table~\ref{TAB:\TABf} of Additional Material.
This emission was not found experimentally,
which was interpreted as due to non-radiative decay to an anomalous state.~\cite{BESSIERE:06}
The lack of this emission 
in Ce-doped \HostCsLiYCl\ was interpreted in the same way~\cite{BESSIERE:04};
in Ce-doped \HostCsNaYCl\ the responsibility of \PrIII\ and/or \NdIII\ killer sites,
which are present in low concentrations and absorb in the same energy region,
was considered instead.~\cite{TANNER:03}
In overall, the calculations of the $5d^1$ manifold
 underestimate the energy of $^2T_{2g}$ and give a good energy of $^2E_g$;
this indicates an overestimation of the crystal-field splitting ($10Dq$) together with an underestimation of the $5d^1$ barycenter.
$^2T_{2g}$ energy underestimations of 2500~\cmm1\ and $^2E_g$ good energies were previously found in 
similar calculation on Ce-doped \HostCsNaYCl\ (Ref.~\onlinecite{TANNER:03}).

\iftoggle{journal-like}{
  \begin{widetext}\begin{center} 
  \begin{table}[h]
\caption{
         Spectroscopic constants of the \CeIII\ and \CeIV\ individual centers in Ce-doped \HostCsLiLuCl: 
         Ce-Cl equilibrium distance \dCeCle,
         breathing mode harmonic vibrational frequency \nue (\cmm1),
         and minimum-to-minimum transition energy \Te. 
         Total energy differences calculated at the equilibrium geometries of
         the lowest states of the $4f^1$, $5dt_{2g}^1$, $5de_{g}^1$, and ITE-$a_{1g}^1$
         configurations of \CeIII, and of the ground state of \CeIV.
         Energies in \cmm1, distances in \AA.
         Identification of energy differences with the absorption and emission transitions of Fig.~\ref{FIG:\FIGa}
         are shwown in parentheses.
              }
\label{TAB:CeIII_res}
\begin{ruledtabular}
\begin{tabular}{ccccc ccccc}
       &       &          &      &     & \multicolumn{5}{c}{\dCeCl\ (\AA)} \\
Branch & State & \dCeCle\ & \nue & \Te & 2.660 & 2.618 & 2.669 & 2.555 & 2.542 \\
\multicolumn{5}{l}{A~~~\CeIII\ $4f^1$}  & &{\bf (2)}&{\bf (4)}&{\bf (6)} \\
  & $1\Gamma_{7u}$($^2A_{2u}$)  & 2.660 & 313 &    0  &      0  & -23340 & -46030 & -52050 & -34880 \\
  & $1\Gamma_{8u}$($^2T_{2u}$)  & 2.662 & 314 &  530  &    540  & -22740 & -45500 & -51360 & -34160 \\
  & $2\Gamma_{7u}$($^2T_{2u}$)  & 2.660 & 314 & 2340  &   2340  & -21010 & -43690 & -49720 & -32550 \\
  & $2\Gamma_{8u}$($^2T_{1u}$)  & 2.662 & 314 & 2800  &   2800  & -20490 & -43230 & -49120 & -31920 \\
  & $1\Gamma_{6u}$($^2T_{1u}$)  & 2.663 & 314 & 3040  &   3040  & -20220 & -43000 & -48800 & -31590 \\
\multicolumn{5}{l}{B~~~\CeIII\ $5dt_{2g}^1$}  & {\bf (1)} \\                                      
  & $1\Gamma_{8g}$($^2T_{2g}$)  & 2.618 & 318 & 23930 &   24500 &      0 & -21290 & -30620 & -13860  \\
  & $1\Gamma_{7g}$($^2T_{2g}$)  & 2.618 & 319 & 25120 &   25710 &   1190 & -20080 & -29460 & -12710  \\
\multicolumn{5}{l}{C~~~\CeIII\ $5de_{g}^1$}  & {\bf (3)} \\                                      
  & $2\Gamma_{8g}$($^2E_{g}$)~\footnotemark[1]
                                & 2.669 & 322 & 45860 &   46070 &  22910 &      0 &  -5780 &  11370  \\
\multicolumn{5}{l}{D~~~\CeIII\ ITE-$a_{1g}^1$}  & {\bf (5)} \\                         
  & $1\Gamma_{6g}$($^2A_{1g}$)  & 2.555 & 325 & 55810 &   59190 &  33220 &  13690 &      0 &  16170  \\
\multicolumn{5}{l}{\CeIV\ + $e^{-}$({\it vacuum})} \\
  & $A_{1g}$($^1A_{1g}$)        & 2.542 & 355 & 39690 &   44870 &  18050 &  -420  & -16050 &      0 \\
\end{tabular}
\end{ruledtabular}
\footnotetext[1]{The calculation of the $E \otimes e_g$ Jahn-Teller coupling of the $2\Gamma_{8g}$($^2E_{g}$) state gives
a $D_{4h}$ equilibrium structure with the following data:
$\delta_{e_g}$=0.081~\AA,
$d_{\rm Ce-Cl,ax}$=2.831~\AA, $d_{\rm Ce-Cl,eq}$=2.588~\AA, $\omega_{e_g}$=228~\cmm1,
$E_{JT}$=1670~\cmm1, T$_e(\Gamma_{6g}(^2A_{1g}))$=44190~\cmm1,
and an energy barrier between equivalent minima of 1200~\cmm1: T$_e(\Gamma_{7g}(^2B_{1g}))$=45390~\cmm1.}

  \end{table}
  \end{center}\end{widetext}
}{}

\iftoggle{journal-like}{
  \def\escalafig{0.45} 
  \begin{figure}
\resizebox{\escalafig\textwidth}{!}{
  \begin{tabular}{c}
    \rotatebox{00}{\includegraphics[scale=1,clip]{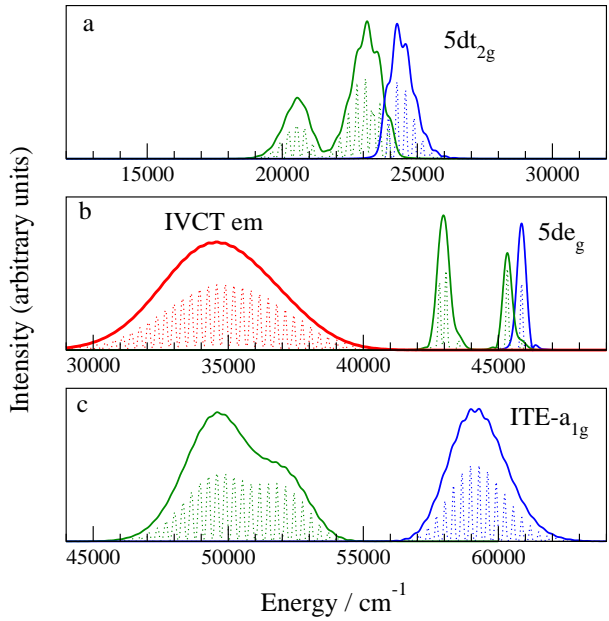}}
  \end{tabular}
}
\caption{
         Simulated absortion and emission band shapes.
}
\label{FIG:abs-em}
  \end{figure}
}{}

A weak broad band in the excitation spectra peaking  at 182~nm (54900~\cmm1) 
was assigned to an impurity-trapped exciton and
next to it, at 176~nm (56800~\cmm1), another more intense but still weak broad band was assigned to free excitons~\cite{BESSIERE:06}
(A7 and A8 in Table~\ref{TAB:\TABI}).
The calculations give a $\Gamma_{6g}$ state of ITE-$a_{1g}^1$ character 55800~\cmm1\ above the ground state,
with the maximum of the absorption band at 59200~\cmm1\
({\bf 5} in Fig.~\ref{FIG:\FIGIV} and Table~\ref{TAB:\TABII}).
This state has an equilibrium Ce-Cl distance of 2.555~\AA, 
which is much shorter than in the $4f^1$, $5dt_{2g}^1$ and $5de_{2g}^1$ states and not far from the 2.542~\AA\
 of the \CeIV\ center. 
Its vibrational frequency (325~\cmm1) is only slighlty larger than in the $4f^1$, $5dt_{2g}^1$ and $5de_{2g}^1$ states
and it does not get as close to the 355~\cmm1\ of the \CeIV\ center.
Its emission is predicted at 48800-52050~\cmm1\ ({\bf 6} in Fig.~\ref{FIG:\FIGIV} and Table~\ref{TAB:\TABII}).
In Fig.~\ref{FIG:\FIGj}c we show  
the simulation of the shape of these absortion (blue lines) and emission (green lines) bands.
The oscillator strengths of Table~\ref{TAB:\TABf} of Additional Material have been used for the emisson band.
The absorption is electric dipole forbidden in $O_h$ symmetry and we used an arbitrary oscillator strength.
In this way we are able to see the band shape, which is made of a very long vibrational progression;
note, however, that the intensities of the absorption and emission band in Fig.~\ref{FIG:\FIGj}c cannot be compared.

The calculations of the \CeIII\ center do not give any state that could be considered responsible for 
the very broad emission at 36400~\cmm1.
This is so in spite of the fact that the
basis sets and spaces used in the calculations are flexible enough to 
reveal the existence
electron-hole electronic structures, like those associated so far with
the excited states responsible for 
anomalous luminescence in lanthanide-doped crystals, should they
occur.\cite{MCCLURE:85,MOINE:89,DORENBOS:03:a}

\subsection{\CeIII--\CeIV\ active pair}
\label{SEC:Pair}

It is not uncommon that \CeIV\ is present in \CeIII-doped solids.~\cite{VANSCHAIK:93}
Having in mind the possibility that this is the case in \CeIII-doped elpasolites,
we discuss in this Section the basics of the electronic structure of the \CeIII--\CeIV\ pairs in Ce-doped \Host.
We show what absorptions and emissions can take place that are associated with these pairs and 
cannot be present in single \CeIII\ active centers. 
In particular, we focus on the intervalence charge transfer luminescence that can be responsible
for the so called anomalous luminescence of this material.

The diabatic potential energy surfaces of the ground and excited states of the \CeIII--\CeIV\ active pair
as functions of the Ce-Cl distances in the left and right CeCl$_6$ moieties
have been calculated using Eq.~\ref{EQ:DiabaticPES}
as described in Sec.~\ref{SEC:diabatic_PES}.
Those of the ground state are shown in Fig.~\ref{FIG:\FIGh}. 
Their main features are summarized in Table~\ref{TAB:\TABc}.
The energies of the states as functions of the ground state electron transfer reaction coordinate are shown in Fig.~\ref{FIG:\FIGc}.
In Fig.~\ref{FIG:\FIGd} of Additional Material
we represent them as functions of the electron transfer reaction coordinates
of excited states.
Let us briefly describe the meaning of the notation in these Figures and in Table~\ref{TAB:\TABc}.

\iftoggle{journal-like}{
  \def\escalafig{0.4} 
  \begin{figure} 
\resizebox{\escalafig\textwidth}{!}{
  \begin{tabular}{c}
  \hspace{18mm}
    \includegraphics[scale=0.25,clip]{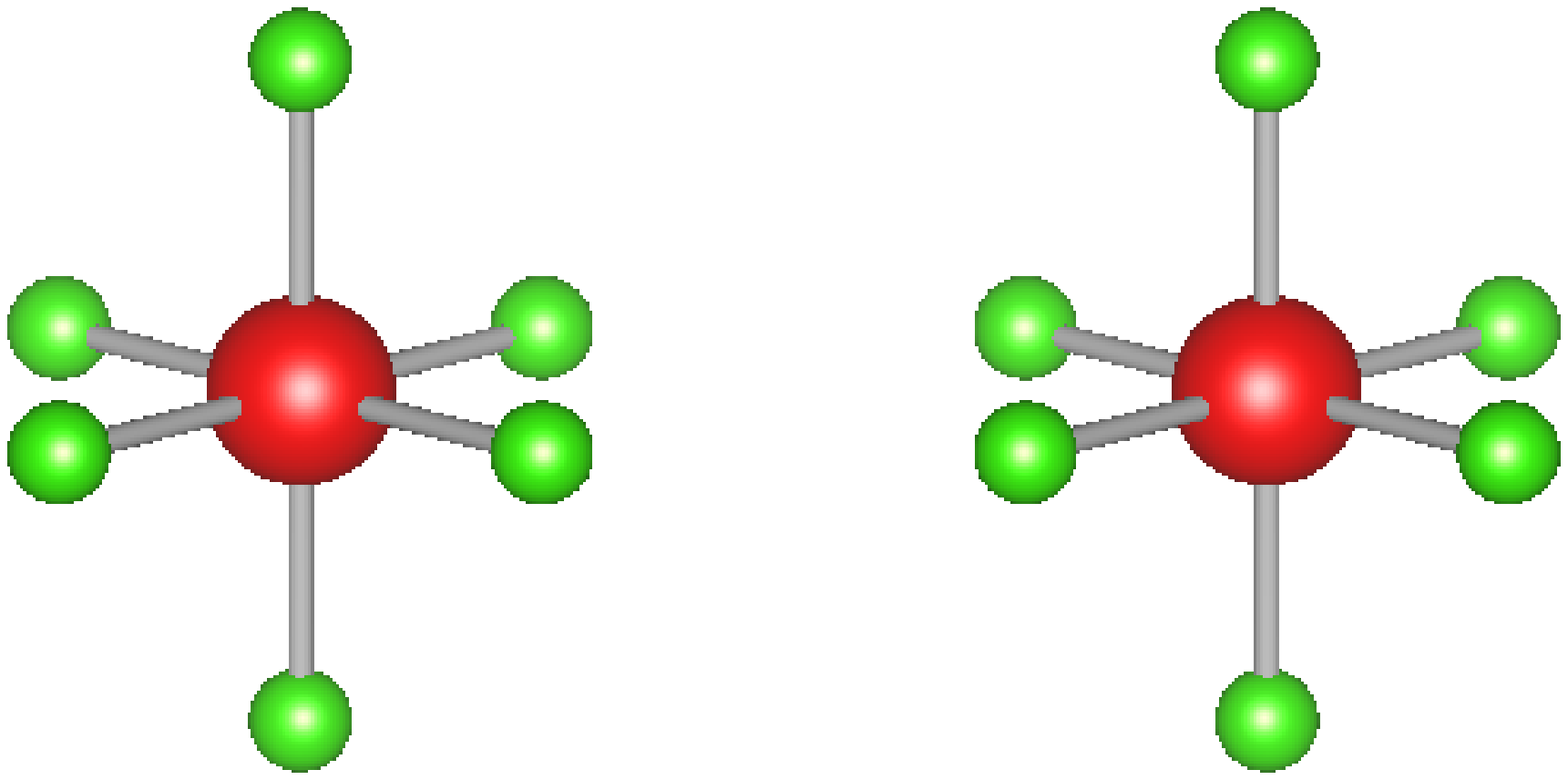}\hspace{-10mm}
\\[-19mm]
    \hspace{29mm}
    \includegraphics[scale=0.25,clip]{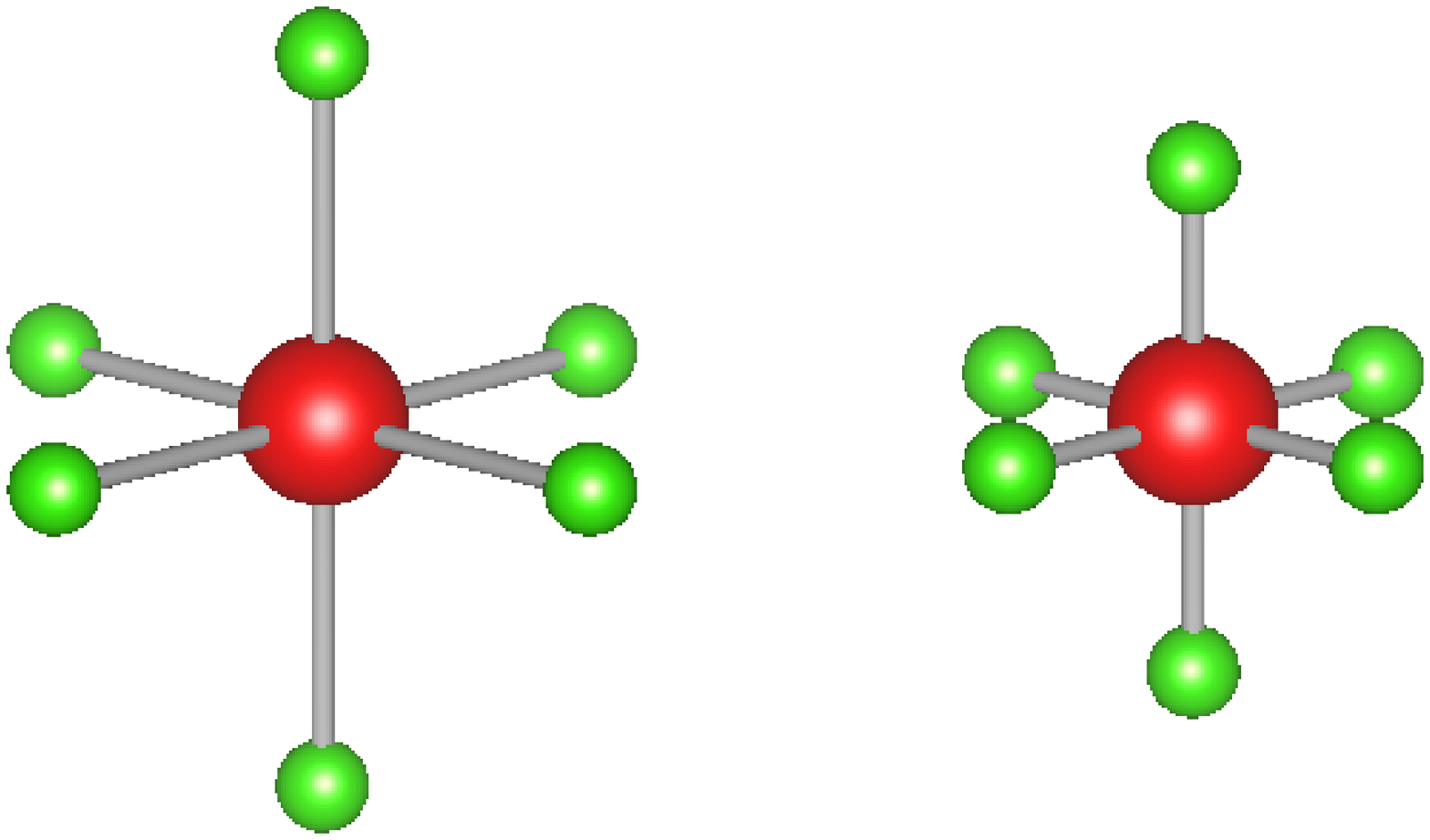}\hspace{30mm}
    \includegraphics[scale=0.25,clip]{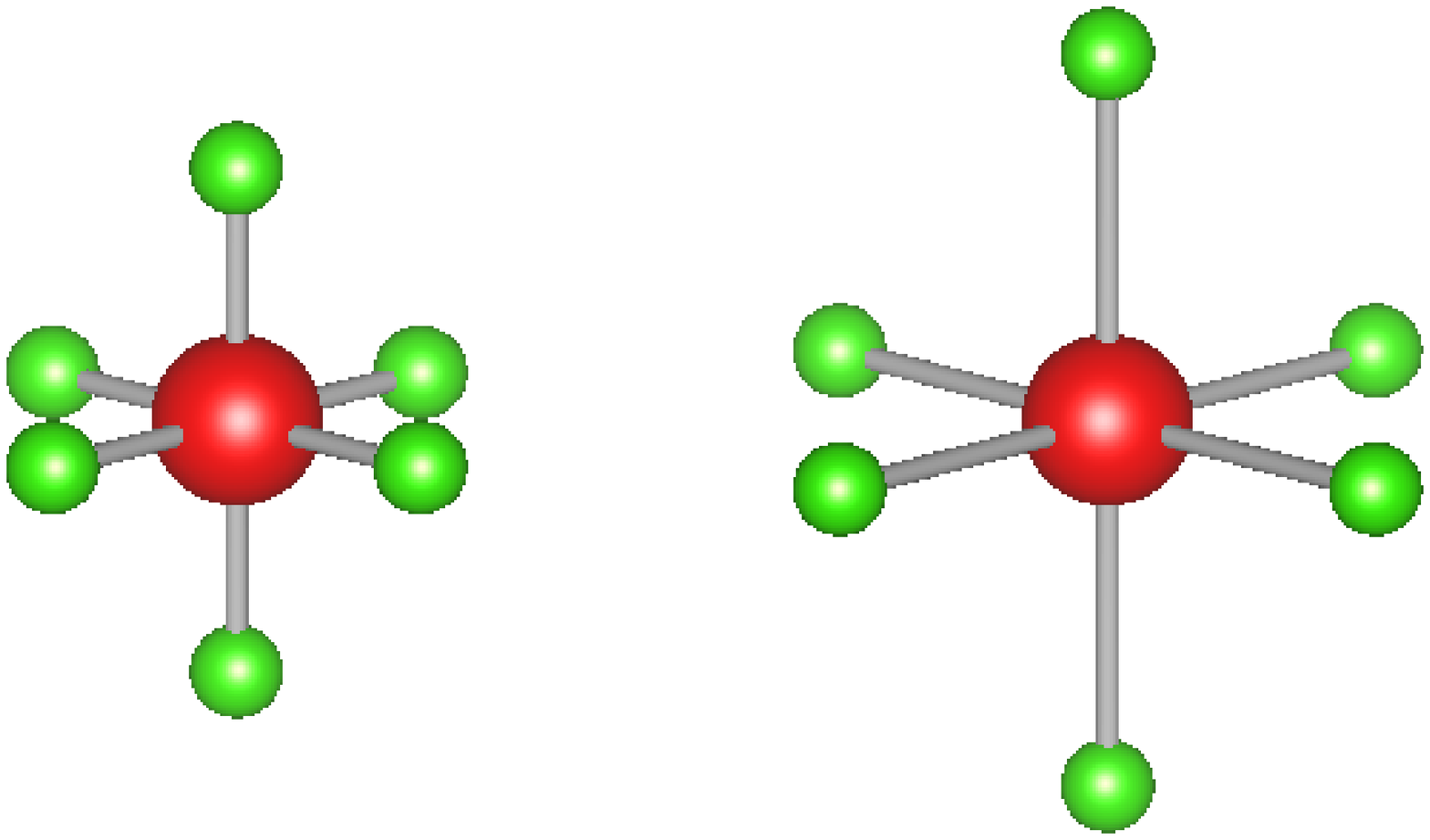}
\\[-10mm]
    \rotatebox{00}{\includegraphics[scale=1,clip]{CeIII-CeIV-11.eps}}
  \end{tabular}
}
\caption{
Diabatic IVCT
         energy diagram for \CeIII-\CeIV\ embedded pairs:
         Total energies of the \CeClvimiii-\CeClvimii\ cluster pair embedded in \HostCsLiLuCl\
         vs. the electron transfer reaction coordinate of the ground state.
         Ce-Cl distances of the left and right CeCl$_6$ moieties, ${\rm d(Ce-Cl)_L}$ and ${\rm d(Ce-Cl)_R}$, 
         are indicated in the lower scale;
         both of them are equal to 2.599~\AA\ in the activated complex.
         A qualitative representation of the moieties in the left and right branches and in the activated complex is shown at the top.
         Energy scale relative to the ground state energy at equilibrium.
         See text for a description of the labels.
}
\label{FIG:CeIII-CeIV-groundstate}
  \end{figure}
}{}

The $A$ brach contains the  [$4f^1$,$A_{1g}$] electronic states of the \CeIII--\CeIV\ pair.
The $A_{et}$ branch contains the states of the \CeIV--\CeIII\ pair that result 
from the $A$ branch after electron transfer from \CeIII\ to \CeIV.
The  thermal conversion of the $A$ states into $A_{et}$ states
are the electron transfer reactions
\begin{eqnarray}
  {\rm Ce}^{3+}(4f^1)_{\rm eq} + {\rm Ce}^{4+}(A_{1g})_{\rm eq} \rightarrow
  {\rm Ce}^{4+}(A_{1g})_{\rm eq}  + {\rm Ce}^{3+}(4f^1)_{\rm eq} 
\,,
\nonumber
\end{eqnarray}
which involve energy barriers.
The diabatic barriers shown in Fig.~\ref{FIG:\FIGc} are upper bounds of the
adiabatic barriers that result after consideration of the electronic couplings.
The $A^\wedge$ branch represents the stressed $A$ states, 
i.e. the  [$4f^1$,$A_{1g}$] states of the \CeIII--\CeIV\ pair 
with structures close to the equilibrium structure of the \CeIV--\CeIII\ pair in [$A_{1g}$,$4f^1$] states.
Equivalently,
the $A_{et}^\wedge$ branch represents stressed states after electron transfer from the $A$ branch,
i.e. the [$A_{1g}$,$4f^1$] states of the \CeIV--\CeIII\ pair
with structures close to the equilibrium structure of the [$4f^1$,$A_{1g}$] states of the \CeIII--\CeIV\ pair.
Going from branch $A^\wedge$ to $A$ is the non-radiative decay of the [$4f^1$,$A_{1g}$] states
to their equilibrium structure, which involves the release of the reorganization energy.
Obviously, this is energetically equivalent to the non-radiative decay of the [$A_{1g}$,$4f^1$] states,
i.e. going from branch $A_{et}^\wedge$ to $A_{et}$.
The vertical transitions from branch $A$ to $A_{et}^\wedge$ are
the intervalence charge transfer absorptions, 
which take place with fixed nuclei positions
({\bf 7} in Fig.~\ref{FIG:\FIGc} and Table~\ref{TAB:\TABc})
\begin{eqnarray}
  {\rm Ce}^{3+}(4f^1)_{\rm eq} + {\rm Ce}^{4+}(A_{1g})_{\rm eq} \stackrel{h\nu}{\rightarrow}
  {\rm Ce}^{4+}(A_{1g})_{\rm str}  + {\rm Ce}^{3+}(4f^1)_{\rm str} 
\nonumber
\,.
\end{eqnarray}
They would inmediatly be followed by the non-radiative decay to $A_{et}$,
which involves the simultaneous relaxation of the coordination shells of \CeIV\ and \CeIII, 
\begin{eqnarray}
  {\rm Ce}^{4+}(A_{1g})_{\rm str}  + {\rm Ce}^{3+}(4f^1)_{\rm str} \leadsto
  {\rm Ce}^{4+}(A_{1g})_{\rm eq}  + {\rm Ce}^{3+}(4f^1)_{\rm eq} 
\nonumber
\,.
\end{eqnarray}

\iftoggle{journal-like}{
  \begin{widetext}\begin{center} 
  \begin{table}[h]
\caption{
         Absorption and emission peak positions of the \CeIII--\CeIV\ pair in \HostCsLiLuCl,
         calculated as total energy differences at the equilibrium geometries of
         the lowest states of the $4f^1$, $5dt_{2g}^1$, $5de_{g}^1$, and ITE-$a_{1g}^1$ 
         configurations of \CeIII\ and of the $A_{1g}$ state of \CeIV.
         \dCeClL, \dCeClR\ and \dCeClAC\ are the Ce-Cl distances in the left and right CeCl$_6$ moieties
         and in the activated complex, respectively.
         Energies in \cmm1, distances in \AA.
         Identification of energy differences with absorption and emission transitions of Fig~\ref{FIG:\FIGc} are shown
         in parentheses.
              }
\label{TAB:CeIII-CeIV}
\begin{ruledtabular}
\begin{tabular}{cccccc}
\multicolumn{6}{c}{Absorption and emission peaks}\\
           &\dCeClL,\dCeClR:
           & 2.660,2.542 
           & 2.618,2.542 
           & 2.669,2.542 
           & 2.555,2.542 \\
          && $4f$ absorption & $5dt_{2g}$ emission & $5de_g$ emission & $a_{1g}$ ITE emission \\
           & {Initial state [\CeIII,\CeIV]:} 
           & $[1\Gamma_{7u}$($^2A_{2u}$),$A_{1g}]$ $\rightarrow$
           & $[1\Gamma_{8g}$($^2T_{2g}$),$A_{1g}]$ $\rightarrow$
           & $[2\Gamma_{8g}$($^2E_{g}$),$A_{1g}]$  $\rightarrow$
           & $[1\Gamma_{6g}$($^2A_{1g}$),$A_{1g}]$ $\rightarrow$ \\
 Branch & Final state \\
&                                                    &       &{\bf (2)}&{\bf (4)}&{\bf (6)}\\
  \multicolumn{1}{c}{$A$} 
& \multicolumn{1}{c}{[\CeIII($4f^1$),\CeIV]}         
    & \CeIII\ $4f         \rightarrow 4f$ 
    & \CeIII\ $5dt_{2g}   \rightarrow 4f$ 
    & \CeIII\ $5de_{g}    \rightarrow 4f$ 
    & \CeIII\ ITE-$a_{1g} \rightarrow 4f$ \\ 
  & $\rightarrow$ $[1\Gamma_{7u}(^2A_{2u}),A_{1g}]$  &    0  &  23340 &  46030 &  52050 \\
  & $\rightarrow$ $[1\Gamma_{8u}(^2T_{2u}),A_{1g}]$  &  540  &  22740 &  45500 &  51360 \\
  & $\rightarrow$ $[2\Gamma_{7u}(^2T_{2u}),A_{1g}]$  & 2340  &  21010 &  43690 &  49720 \\
  & $\rightarrow$ $[2\Gamma_{8u}(^2T_{1u}),A_{1g}]$  & 2800  &  20490 &  43230 &  49120 \\
  & $\rightarrow$ $[1\Gamma_{6u}(^2T_{1u}),A_{1g}]$  & 3040  &  20220 &  43000 &  48800 \\
&                                                    &{\bf (7)}&      &{\bf (10)}&{\bf (11)} \\
  \multicolumn{1}{c}{$A_{\rm et}^\wedge$ } 
& \multicolumn{1}{c}{[\CeIV,\CeIII($4f^1$)]$_{\rm sts}$} 
    &\CeIII $4f$         $\rightarrow$ \CeIV $4f$
    &\CeIII $5dt_{2g}$   $\rightarrow$ \CeIV $4f$
    &\CeIII $5de_g$      $\rightarrow$ \CeIV $4f$
    &\CeIII ITE-$a_{1g}$ $\rightarrow$ \CeIV $4f$ \\
  & $\rightarrow$ $[A_{1g},1\Gamma_{7u}(^2A_{2u})]$  & 10000 &  16820 &  35290 &  50920  \\
  & $\rightarrow$ $[A_{1g},1\Gamma_{8u}(^2T_{2u})]$  & 10720 &  16100 &  34570 &  50200  \\
  & $\rightarrow$ $[A_{1g},2\Gamma_{7u}(^2T_{2u})]$  & 12330 &  14490 &  32960 &  48590 \\
  & $\rightarrow$ $[A_{1g},2\Gamma_{8u}(^2T_{1u})]$  & 12950 &  13870 &  32340 &  47970 \\
  & $\rightarrow$ $[A_{1g},1\Gamma_{6u}(^2T_{1u})]$  & 13290 &  13530 &  32000 &  47630 \\
&                                                  &{\bf (1)}&        &        &        \\
  \multicolumn{1}{c}{$B$ } 
& \multicolumn{1}{c}{[\CeIII($5dt_{2g}^1$),\CeIV]} 
    &\CeIII $4f         \rightarrow 5dt_{2g}$
    &        
    &        
    &        \\
  & $\rightarrow$ $[1\Gamma_{8g}(^2T_{2g}),A_{1g}]$  & 24500 &      0 &  21290 &  30620 \\
  & $\rightarrow$ $[1\Gamma_{7g}(^2T_{2g}),A_{1g}]$  & 25710 &        &  20080 &  29460  \\
&                                                              & {\bf (8)} \\
  \multicolumn{1}{c}{$B_{\rm et}^\wedge$ } 
& \multicolumn{1}{c}{[\CeIV,\CeIII($5dt_{2g}^1$)]$_{\rm sts}$} 
    &\CeIII $4f$         $\rightarrow$ \CeIV $5dt_{2g}$
    &        
    &        
    &       \\
  & $\rightarrow$ $[A_{1g},1\Gamma_{8g}(^2T_{2g})]$  & 31010 &        &  14280 &  29910 \\
  & $\rightarrow$ $[A_{1g},1\Gamma_{7g}(^2T_{2g})]$  & 32170 &        &  13120 &  28750 \\
&                                                  &{\bf (3)}&        &        &       \\
  \multicolumn{1}{c}{$C$ } 
& \multicolumn{1}{c}{[\CeIII($5de_{g}^1$),\CeIV]}  
    &\CeIII $4f         \rightarrow 5de_g$ 
    &        
    &        
    &       \\
  & $\rightarrow$ $[2\Gamma_{8g}(^2E_{g}),A_{1g}]$   & 46070 &        &      0 &  5780 \\
&                                                              &{\bf (9)} \\
  \multicolumn{1}{c}{$C_{\rm et}^\wedge$ } 
& \multicolumn{1}{c}{[\CeIV,\CeIII($5de_{g}^1$)]$_{\rm sts}$}  
    &\CeIII $4f$         $\rightarrow$ \CeIV $5de_g$
    &        
    &        
    &       \\
  & $\rightarrow$ $[A_{1g},2\Gamma_{8g}(^2E_{g})]$   & 56240 &        &        &  4680 \\
&                                                      &{\bf (5)}&        &        &       \\
  \multicolumn{1}{c}{$D$ } 
& \multicolumn{1}{c}{[\CeIII(ITE-$a_{1g}\,^1$),\CeIV]} 
    &\CeIII $4f         \rightarrow$ ITE-$a_{1g}$ 
    &        
    &        
    &       \\
  & $\rightarrow$ $[1\Gamma_{6g}(^2A_{1g}),A_{1g}]$  & 59190 &        &        & 0 \\
  \multicolumn{1}{c}{$D_{\rm et}^\wedge$ } 
& \multicolumn{1}{c}{[\CeIV,\CeIII(ITE-$a_{1g}\,^1$)]$_{\rm sts}$} 
    &
    &        
    &        
    &       \\
  & $\rightarrow$ $[A_{1g},1\Gamma_{6g}(^2A_{1g})]$  & 61050 &        &        &  \\
\hline
\multicolumn{6}{c}{Branch crossings}\\
\multicolumn{2}{c}{Branch pair and energy barrier} 
                           & $A$-$A_{\rm et}$ 2600 & $B$-$B_{\rm et}$ 1100 & $C$-$C_{\rm et}$ 2900 & $D$-$D_{\rm et}$ 50 \\
\multicolumn{2}{c}{\dCeClAC} &        2.599 &                 2.578 &                 2.600 &               2.548 \\
\multicolumn{2}{c}{Branch pair and energy barrier}
                            & $B$-$A_{\rm et}^\wedge$ 5800 & $C$-$B_{\rm et}^\wedge$ 5100 & $D$-$C_{\rm et}^\wedge$ 1100 \\
\multicolumn{2}{c}{\dCeClL, \dCeClR}       & 2.70, 2.45               & 2.79, 2.49               & 2.56, 2.49 \\
\end{tabular}
\end{ruledtabular}
  \end{table}
  \end{center}\end{widetext}
}{}

The $B$, $C$, and $D$ related branches are equivalent to the $A$ ones,
but referred to the states of \CeIII\ of the $5dt_{2g}^1$ configuration,
the $5de_{g}^1$ configuration, and the ITE-$a_{1g}^1$ impurity trapped excitonic character, respectively.

In Fig.~\ref{FIG:\FIGc} and Table~\ref{TAB:\TABc}
we have indicated the photo-processes that take place in the \CeIII\ active center
and are responsible for observed spectroscopic features, as discussed in the previous section:
The \mbox{$4f$ $\rightarrow$ $5dt_{2g}$}, 
    \mbox{$4f$ $\rightarrow$ $5de_{g}$}, and 
    \mbox{$4f$ $\rightarrow$ ITE-$a_{1g}$} absorptions ({\bf 1, 3, 5}), and 
the \mbox{$5dt_{2g}$ $\rightarrow$ $4f$} emission ({\bf 2}).

\subsubsection{IVCT absorptions}

More interesting now
are the photo-processes that involve intervalence charge transfer between \CeIII\ and \CeIV.
In Fig.~\ref{FIG:\FIGc} they correspond to vertical lines starting in the minimum of one branch
($A$, $A_{et}$, $B$, $B_{et}$, $C$, $C_{et}$, $D$, $D_{et}$),
drawn with full lines,
and ending in one of the stressed branches
($A^\wedge$, $A_{et}^\wedge$, $B^\wedge$, $B_{et}^\wedge$, $C^\wedge$, $C_{et}^\wedge$, $D$, $D_{et}^\wedge$),
drawn with dashed lines.
We have indicated three of them in Fig.~\ref{FIG:\FIGc}.
All of these transitions give very broad bands because of the large offsets between the minima
of their corresponding energy surfaces: $A$ and $A_{et}$, $A$ and $B_{et}$, etc.
These offsets are given by $\Delta Q = Q_{e,f} - Q_{e,i}$,
with the equilibrium coordinates of the final and initial state  $Q_{e,f}$ and  $Q_{e,i}$ given by Eq.~\ref{EQ:Qet}.
The values of these horizontal offsets are given in Table~\ref{TAB:\TABg} of Additional Material.
They are approximately equal to $\sqrt{3} (\Delta d_R - \Delta d_L)$,
$\Delta d_R$ and $\Delta d_L$ being the Ce-Cl bond length changes experienced by the \CeIV\ acceptor center (right)
and the \CeIII\ donor center (left) in the IVCT transition,
which, in turn, is about $2\sqrt{3}$ times the difference of Ce-Cl equilibrium distances in
the \CeIII\ and \CeIV\ centers.

Absorption {\bf 7}, 
$A \rightarrow A_{et}^\wedge$, or
\mbox{[\CeIII($4f^1$),\CeIV]$_{\rm eq}$} $\rightarrow$ \mbox{[\CeIV,\CeIII($4f^1$)]$_{\rm sts}$},
is the photoinduced electron transfer 
commonly found in mixed valence compounds,~\cite{WERNER:96,ALLEN:67,ROBIN:68}
i.e. the IVCT absorption.
It has been reported in \CeIII-doped LaPO$_4$ by Van Schaik \etal~\cite{VANSCHAIK:93}
It corresponds to the vertical one-electron orbital transition \mbox{\CeIII $4f$} $\rightarrow$ \mbox{\CeIV $4f$}
and it is followed by a strong nuclei reorganization around \CeIV\ and \CeIII\
in which all the IVCT absortion energy is released:
\mbox{[\CeIV,\CeIII($4f^1$)]$_{\rm sts}$} $\leadsto$ \mbox{[\CeIV,\CeIII($4f^1$)]$_{\rm eq}$},
The present calculations predict it for the \CeIII--\CeIV\ pair in \Host\
as a very broad band peaking at about 10000-13300~\cmm1,
which can also occur in other chloroelpasolites. 
The large width of this band is associated
with the large offset between the minima of the $A$ and $A_{et}$ energy surfaces,
which is here found to be of about 0.41~\AA\ (cf. Table~\ref{TAB:\TABg} of Additional Material).

Absorption {\bf 8},  
$A \rightarrow B_{et}^\wedge$,  or 
\mbox{[\CeIII($4f^1$),\CeIV]$_{\rm eq}$} $\rightarrow$ \mbox{[\CeIV,\CeIII($5dt_{2g}^1$)]$_{\rm sts}$},
is an IVCT absorption of higher energy.
It corresponds to the vertical one-electron orbital transition \mbox{\CeIII $4f$} $\rightarrow$ \mbox{\CeIV $5dt_{2g}$}.
Its energy is above the regular \CeIII\ \mbox{$4f$ $\rightarrow$ $5dt_{2g}$} absorption ({\bf 1}) and
the difference between them is the reorganization energy that is released after the absorption.
In this case, it is predicted as a very broad band peaking at 31000-32200~\cmm1,
which is 6500~\cmm1\ above \CeIII\ \mbox{$4f$ $\rightarrow$ $5dt_{2g}$}.
Since the latter is underestimated in these calculations, we can expect the same degree of underestimation in 
this IVCT absortion.
The experimental peaks in the excitation spectrum at 33000~\cmm1\ (A3) and 35700~\cmm1\ (A4), which are very wide and
have been unassigned (A3) or assigned to lower symmetry \CeIII\ sites~\cite{BESSIERE:06} (A4)
can be assigned to this IVCT absorption.
Again, the large width of this band is associated
with the large offset between the minima of the $A$ and $B_{et}$ energy surfaces,
which is here found to be of about 0.34~\AA\ (cf. Table~\ref{TAB:\TABg} of Additional Material).

Absorption {\bf 9},
$A \rightarrow C_{et}^\wedge$,  or 
\mbox{[\CeIII($4f^1$),\CeIV]$_{\rm eq}$} $\rightarrow$ \mbox{[\CeIV,\CeIII($5de_g^1$)]$_{\rm sts}$}, 
is an IVCT absorption of even higher energy.
Analogously to absorption 8, it is
the vertical one-electron orbital transition \mbox{\CeIII $4f$} $\rightarrow$ \mbox{\CeIV $5de_g$}
and
its energy is above the regular \CeIII\ \mbox{$4f$ $\rightarrow$ $5de_g$} absorption ({\bf 3})
in an amount equal to the reorganization energy.
It is predicted as a very broad band peaking at 56240~\cmm1,
10200~\cmm1\ above \CeIII\ \mbox{$4f$ $\rightarrow$ $5de_g$}.
Since the latter is calculated to be close to experiment in these calculations, we can expect analogous agreement in 
this IVCT absortion.
The experimental peak in the excitation spectrum at 54900~\cmm1\ (A7), can be associated with this IVCT absorption. 
The fact that the intensity of the A7 band in the excitation spectrum of the $dt_{2g}$ emission is comparable to that of the
\CeIII\ \mbox{$4f$ $\rightarrow$ $5de_g$} (A2) band further suggests this assignement.

\subsubsection{IVCT luminescence}

Emission {\bf 10}, 
$C \rightarrow A_{et}^\wedge$,  or 
\mbox{[\CeIII($5de_{g}^1$),\CeIV]$_{\rm eq}$} $\rightarrow$ \mbox{[\CeIV,\CeIII($4f^1$)]$_{\rm sts}$},
is specially interesting because it is an IVCT emission and,
to the best of our knowledge, IVCT emissions have never been reported.
As we discuss next, 
this emission corresponds well with the so called anomalous emission (E2).~\cite{BESSIERE:06}
This IVCT luminescence 
is the vertical one-electron orbital transition \mbox{\CeIII $5de_g$} $\rightarrow$ \mbox{\CeIV $4f$}
and it is followed by a strong nuclei reorganization around \CeIV\ and \CeIII.
This involves a large Ce-Cl distance shortening around  \CeIV\ and a large Ce-Cl distance elongation around \CeIII,
and it releases an energy which is very similar to the energy of the IVCT absorption {\bf 7}.
The only difference between 
these energies
arises from the fact that the Ce-Cl equilibrium bond length in the \CeIII\ center
is different in its $5de_g^1$ states and $4f^1$ states,
but this difference is much smaller than the offset between the minima of the
$C$ and $A_{et}$ branches along the electron transfer normal coordinate $Q_{eg}$,
which determines the reorganization energy.
The reorganization energy  after the IVCT emission is the amount by which this emission is lower than 
the \CeIII\ \mbox{$4f$ $\rightarrow$ $5de_{g}$} absorption, 
i.e. it is its Stokes shift.
In other words, the Stokes shift of this IVCT emission amounts the ground IVCT absorption {\bf 7}.
This is slightly above 10000~\cmm1\ in the present calculations.
The experimental Stokes shift of the anomalous emission, calculated as
the difference between the experimental \mbox{$4f$ $\rightarrow$ $5de_{g}$} absorption maximum
(46900~\cmm1\ average of the JT split excitation)
and the anomalous emission maximum (36400~\cmm1) is 10500~\cmm1.

The calculation predicts this IVCT emission band to be a very wide band
with peaks in the 32000-35300~\cmm1\ region (Table~\ref{TAB:\TABc}).
Its simulation is shown in red in Fig.~\ref{FIG:\FIGj}b.
Since the diabatic calculations do not provide oscillator strengths, we have used
the oscillator strengths of the regular $5de_g \rightarrow 4f$ emission in Table~\ref{TAB:\TABf} of Additional Material
in order to be able to simulate the band shape. 
Accordingly, the intensities of the IVCT emission band and the $5de_g \rightarrow 4f$ emission band in Fig.~\ref{FIG:\FIGj}b
cannot be compared, but we  must expect that the different shapes of both emission bands are fairly well reproduced.
The IVCT emission is
 a very broad band with a full width at half maximum (FWHM) of about 5000~\cmm1.
The experimental FWHM at 290~K is 4800~\cmm1.
So, both the position and shape of this band correspond well with the anomalous emission (E2).
Moreover,
the following additional supporting argument on the band width can be given:
First,
we should expect the IVCT emission {\bf 10} to have a band shape very similar to that of
the regular IVCT absorption {\bf 7},
because both of them have very similar normal coordinate offsets.
Second, the band width $\Delta\bar\nu_{1/2}$ (in \cmm1) and band maximum $\bar\nu_{max}$ (in \cmm1) 
of the latter are related according to
$\Delta\bar\nu_{1/2} = (\bar\nu_{max}\,7.71\,T)^{1/2}$  
at temperatures T (in K) high enough so that $\hbar\omega << 2kT$, $\omega$ being the vibrational frequency;~\cite{HUSH:67}
and using the experimental Stokes shift of the anomalous emission for $\bar\nu_{max}$, $\bar\nu_{max}$=10500~\cmm1,
we get at T=290~K a FWHM of $\Delta\bar\nu_{1/2}$=4850~\cmm1,
which agrees very well with the experimental FWHM of the anomalous emission.

According to the interpretation given in this paper,
the anomalous emission is not due to an electron-hole recombination in which the electron is in the conduction band and the hole in
the Ce impurity, and there is no anomalous state. 
Instead, the emission is an electron transfer from a $5de_g$ orbital of a \CeIII\ center
to a $4f$ orbital of a \CeIV\ center, which takes place with frozen nuclei positions
and leaves the newly formed \CeIV\ and \CeIII\ centers in their ground electronic states and 
under strong structural stress. 
The large relaxation energy released afterwars is basically the large Stokes shift of this emission. 
This is graphically represented in Fig.~\ref{FIG:\FIGf} of Additional Material. 
If an electron-hole recombination description is to be used, the electron is in a \CeIII\ center and the hole in a \CeIV\ center,
but this kind of description is not very adequate because such a recombanition of electron and hole 
would still leave a hole and an electron.

The principal excitation of the anomalous emission is the \CeIII\ \mbox{$4f$ $\rightarrow$ $5de_{g}$} absorption
(A2, cf. Figs.~5c and 5d in Ref.~\onlinecite{BESSIERE:06});
this is in agreement with the emission starting in the $5de_{g}^1$ manifold.
Also, two additional broad peaks have been identified in its excitation spectrum at 54300~\cmm1\ (A9) and 57500~\cmm1\ (A10);
they can be made of absorptions from the $A$ branch to the $C_{\rm et}^\wedge$, $D$, and $D_{\rm et}^\wedge$ branches,
i.e. to \mbox{\CeIII $4f$} $\rightarrow$ \mbox{\CeIV $5de_g$} IVCT absorption
and absorptions to the impurity-trapped exciton.
We must note that the electronic couplings, which are missing in the diabatic energy surfaces,
will have a stronger effect on the states of these excited branches.
This and the lower symmetry driven splittings also missing here,
indicate that the states at about 5000-10000~\cmm1\ above the \CeIII\ $5de_g^1$ manifold 
should be a strong mix of the diabatic states of the $C_{\rm et}^\wedge$, $D$, and $D_{\rm et}^\wedge$ branches.
They can be responsible for the above mentioned peaks, but also for the peaks at 54900 and 56800~\cmm1\
of the excitation spectrum of the \CeIII\ \mbox{$5dt_{2g}$ $\rightarrow$ $4f$} emission (A7, A8).
The 56800~\cmm1\ peak was assigned to an impurity-trapped exciton in Ref.~\onlinecite{BESSIERE:06}.

Another specific feature of the anomalous emission of Ce-doped \Host\
is that it is not excited by absorptions to the conduction band (CB),
although these absorptions effectively excite the \mbox{$5dt_{2g}$ $\rightarrow$ $4f$} emission.~\cite{BESSIERE:06}
Such absortions mean an ionization of \CeIII\ to the CB that destroys the \CeIII--\CeIV\ pair, so that
for the anomalous emission to happen, the electron must be transferred back from the CB  to the $5de_g$ empty orbital of \CeIV.
The present calculation does not give information on this;
calculations on  metal-to-metal charge transfer in Ce-Lu, Ce-Cs, and Ce-Li pairs will be the subject of further study.

Finally, in Ref.~\onlinecite{BESSIERE:06} it was shown that
the decay time of the anomalous emission excited with the \mbox{$4f$ $\rightarrow$ $5de_{g}$} absorption
was found to remain constant from 12~K to 240~K, 
to start dropping at 240~K (cf. Fig.~8 in Ref.~\onlinecite{BESSIERE:06}),
and to quench at 370~K.
Following its correlation with the \mbox{\temission} emission
it was also shown that the anomalous emission is quenched by means of energy transfer to the $5dt_{2g}^1$ states of \CeIII.
Fitting the parameters of a kinetic model to the experimental data led them to an activation energy of 2900~\cmm1.
Parallely, the \mbox{$4f$ $\rightarrow$ $5de_g$} absorption was found to excite the
\mbox{\temission} emission, although no connection with the above activation energy was made.

All these observations are consistent with the present interpretation. 
In effect,
 the $C$-$B_{et}^\wedge$ crossing is responsible for the decay from the state that emits the anomalous luminescence
(though it is a $5de_g^1$ state and not an anomalous state)
to the $5dt_{2g}^1$ state.
Such crossing gives the activation energy of the thermal process
\mbox{\CeIII($5de_g^1$) + \CeIV} $\rightarrow$ \mbox{\CeIV + \CeIII($5dt_{2g}^1$)}. 
The diabatic result found here is 5100~\cmm1.
This value is reduced by the Jahn-Teller splitting of the $5de_g^1$ state,
it will be further reduced by the inclusion of electronic couplings and,
perhaps more importantly, by a lower crystal-field splitting and
by a larger offset between the minima of the $C$ and $B_{et}$ branches.
We already commented above that the present calculation overestimates the crystal field splitting
and underestimations of the nuclear offsets are common in calculations of this type.~\cite{GRACIA:08}
And this interpretation also explains that the \mbox{$4f$ $\rightarrow$ $5de_g$} absorption excites
the \mbox{$5dt_{2g}$ $\rightarrow$ $4f$} emission.

Besides, 
the quenching of the \mbox{\temission} emission can be related to the $B$-$A_{et}^\wedge$
crossing,
i.e. to the thermal process \mbox{\CeIII($5dt_{2g}^1$) + \CeIV} $\rightarrow$ \mbox{\CeIV + \CeIII($4f^1$)}. 
This is found to have a diabatic activation energy of 5800~\cmm1,
higher than that of the $5de_g^1$ decay but not much. With the same arguments than before, we should expect this barrier to
be reduced upon inclusion of the missing contributions, but not more than the other, so that the results indicate
that the activation energy is slightly larger for the non-radiative decay of the $5dt_{2g}^1$ states 
than of the $5de_g^1$ states. Hence a larger quenching temperature of the \mbox{\temission} emission
compared with the anomalous emission.

Before ending this discussion we may remark that, according to these results,
we may expect the existence of an IVCT emission of lower energy
corresponding to the vertical one-electron orbital transition \mbox{\CeIII $5dt_{2g}$} $\rightarrow$ \mbox{\CeIV $4f$},
which is to say to the transiton 
$B \rightarrow A_{et}^\wedge$,  or 
\mbox{[\CeIII($5dt_{2g}^1$),\CeIV]$_{\rm eq}$} $\rightarrow$ \mbox{[\CeIV,\CeIII($4f^1$)]$_{\rm sts}$}.
This transition is expected to appear at an energy lower than the \mbox{\temission} emission
by an amount very similar the first IVCT absortion, 10000~\cmm1\ in our case, i.e. at about 14000-16000~\cmm1.

The new interpretation of the electronic excited states of Ce-doped \HostCsLiLuCl\ is summarized in 
Table~\ref{TAB:\TABe}.

\section{Conclusions}
\label{SEC:conclusions}

In this paper we report for the first time the existence of intervalence charge transfer luminescence.
We have shown that the so called anomalous luminescence of Ce-doped \HostCsLiLuCl,
which is characterized mainly by a very large Stokes shift and a very large band width,
corresponds to an IVCT emission that takes place in \CeIII--\CeIV\ pairs,
in particular to the vertical one-electron orbital transition \mbox{\CeIII $5de_g$} $\rightarrow$ \mbox{\CeIV $4f$}.
The Stokes shift is the sum of the large reorganization energies of the \CeIV\ and \CeIII\ centers
formed after the electron transfer.
It equals the energy of the IVCT absortion, which is predicted to exist in this material 
and to be slightly larger than 10000~\cmm1.
The large band width is the consequence of the large offset between the minima of the \CeIII--\CeIV\ and
\CeIV--\CeIII\ pairs along the electron transfer reaction coordinate,
which is approximately $2\sqrt{3}$ times the difference of Ce-Cl equilibrium distances in
the \CeIII\ and \CeIV\ centers.

We have shown that the energies of the peaks and the widths of IVCT absortion and emission bands
can be calculated \abinitio\ with reasonable accuracy 
from diabatic energy surfaces of the ground and excited states and that 
these can be obtained, in turn, from independent \abinitio\ calculations on the donor and acceptor active centers.

We obtained the energies of the \CeIII\ and \CeIV\ active centers of Ce-doped \HostCsLiLuCl\
by means of state-of-the-art SA-CASSCF/MS-CASPT2/RASSISO \abinitio\ spin-orbit coupling DKH relativistic calculations 
on the donor cluster \Donor\ and the acceptor cluster \Acceptor\ embedded in an AIMP quantum mechanical 
embedding potential of the host elpasolite \HostCsLiLuCl.
The calculations provide interpretations of  unexplained experimental observations in Ce-doped \HostCsLiLuCl\
as due to higher energy IVCT absorptions. They also allow to reinterpret other observations.
The existence of 
another IVCT emission of lower energy, at around 14000-16000~\cmm1\ less than the 
\mbox{\temission} emission, is also predicted.

IVCT emissions and high energy IVCT absorptions like the ones reported here for \CeIII-\CeIV\ pairs,
which are an extension of the known IVCT absorption of mixed-valence compounds,
have also been found in Yb-doped solids and are reported separately.
They are very likely to exist also in Eu-doped solids
and in solids with f-elements in which several valence states can coexist.

\acknowledgments
This work was partly supported by
a grant from Ministerio de Econom\'{\i}a y Competitivad, Spain
(Direcci\'on General de Investigaci\'on y Gesti\'on del Plan Nacional de I+D+I,
MAT2011-24586).
%

\begin{thebibliography}{56}
\expandafter\ifx\csname natexlab\endcsname\relax\def\natexlab#1{#1}\fi
\expandafter\ifx\csname bibnamefont\endcsname\relax
  \def\bibnamefont#1{#1}\fi
\expandafter\ifx\csname bibfnamefont\endcsname\relax
  \def\bibfnamefont#1{#1}\fi
\expandafter\ifx\csname citenamefont\endcsname\relax
  \def\citenamefont#1{#1}\fi
\expandafter\ifx\csname url\endcsname\relax
  \def\url#1{\texttt{#1}}\fi
\expandafter\ifx\csname urlprefix\endcsname\relax\def\urlprefix{URL }\fi
\providecommand{\bibinfo}[2]{#2}
\providecommand{\eprint}[2][]{\url{#2}}

\bibitem[{\citenamefont{Verhoeven}(1996)}]{VERHOEVEN:96}
\bibinfo{author}{\bibfnamefont{J.~W.} \bibnamefont{Verhoeven}},
  \bibinfo{journal}{Pure Appl. Chem.} \textbf{\bibinfo{volume}{68}},
  \bibinfo{pages}{2223} (\bibinfo{year}{1996}).

\bibitem[{IVC()}]{IVCT-MMCT}
\bibinfo{note}{The term IVCT is used sometimes for MMCT between non-equivalent
  metal ions and also for electron transfer processes not involving metals;
  here we will adopt the conventional meaning and we will call IVCT only to the
  homonuclear, symmetric MMCT.}

\bibitem[{\citenamefont{Marcus}(1964)}]{MARCUS:64}
\bibinfo{author}{\bibfnamefont{R.~A.} \bibnamefont{Marcus}},
  \bibinfo{journal}{Annu. Rev. Phys. Chem} \textbf{\bibinfo{volume}{15}},
  \bibinfo{pages}{155} (\bibinfo{year}{1964}).

\bibitem[{\citenamefont{Allen and Hush}(1967)}]{ALLEN:67}
\bibinfo{author}{\bibfnamefont{G.~C.} \bibnamefont{Allen}} \bibnamefont{and}
  \bibinfo{author}{\bibfnamefont{N.~S.} \bibnamefont{Hush}},
  \bibinfo{journal}{Prog. Inorg. Chem.} \textbf{\bibinfo{volume}{8}},
  \bibinfo{pages}{357} (\bibinfo{year}{1967}).

\bibitem[{\citenamefont{Robin and Day}(1968)}]{ROBIN:68}
\bibinfo{author}{\bibfnamefont{M.}~\bibnamefont{Robin}} \bibnamefont{and}
  \bibinfo{author}{\bibfnamefont{P.}~\bibnamefont{Day}}, \bibinfo{journal}{Adv.
  Inorg. Chem. Radiochem.} \textbf{\bibinfo{volume}{10}}, \bibinfo{pages}{247}
  (\bibinfo{year}{1968}).

\bibitem[{\citenamefont{Werner}(1896)}]{WERNER:96}
\bibinfo{author}{\bibfnamefont{A.}~\bibnamefont{Werner}}, \bibinfo{journal}{Z.
  Anorg. Chem.} \textbf{\bibinfo{volume}{12}}, \bibinfo{pages}{46}
  (\bibinfo{year}{1896}).

\bibitem[{\citenamefont{Hush}(1967)}]{HUSH:67}
\bibinfo{author}{\bibfnamefont{N.~S.} \bibnamefont{Hush}},
  \bibinfo{journal}{Prog. Inorg. Chem.} \textbf{\bibinfo{volume}{8}},
  \bibinfo{pages}{391} (\bibinfo{year}{1967}).

\bibitem[{\citenamefont{Hush}(1985)}]{HUSH:85}
\bibinfo{author}{\bibfnamefont{N.~S.} \bibnamefont{Hush}},
  \bibinfo{journal}{Coord. Chem. Rev.} \textbf{\bibinfo{volume}{64}},
  \bibinfo{pages}{135} (\bibinfo{year}{1985}).

\bibitem[{\citenamefont{Piepho et~al.}(1975)\citenamefont{Piepho, Krausz, and
  Schatz}}]{PIEPHO:78}
\bibinfo{author}{\bibfnamefont{S.~B.} \bibnamefont{Piepho}},
  \bibinfo{author}{\bibfnamefont{E.~R.} \bibnamefont{Krausz}},
  \bibnamefont{and} \bibinfo{author}{\bibfnamefont{P.~N.}
  \bibnamefont{Schatz}}, \bibinfo{journal}{J.\ Amer.\ Chem.\ Soc.}
  \textbf{\bibinfo{volume}{100}}, \bibinfo{pages}{2996} (\bibinfo{year}{1975}).

\bibitem[{\citenamefont{Blasse}(1991)}]{BLASSE:91}
\bibinfo{author}{\bibfnamefont{G.}~\bibnamefont{Blasse}},
  \bibinfo{journal}{Struct. Bond.} \textbf{\bibinfo{volume}{76}},
  \bibinfo{pages}{153} (\bibinfo{year}{1991}).

\bibitem[{\citenamefont{Boutinaud et~al.}(2007)\citenamefont{Boutinaud, Mahiou,
  Cavalli, and Bettinelli}}]{BOUTINAUD:07}
\bibinfo{author}{\bibfnamefont{P.}~\bibnamefont{Boutinaud}},
  \bibinfo{author}{\bibfnamefont{R.}~\bibnamefont{Mahiou}},
  \bibinfo{author}{\bibfnamefont{E.}~\bibnamefont{Cavalli}}, \bibnamefont{and}
  \bibinfo{author}{\bibfnamefont{M.}~\bibnamefont{Bettinelli}},
  \bibinfo{journal}{J.\ Lumin.} \textbf{\bibinfo{volume}{122-123}},
  \bibinfo{pages}{430} (\bibinfo{year}{2007}).

\bibitem[{\citenamefont{{van Schaik} et~al.}(1993)\citenamefont{{van Schaik},
  Lizzo, Smit, and Blasse}}]{VANSCHAIK:93}
\bibinfo{author}{\bibfnamefont{W.}~\bibnamefont{{van Schaik}}},
  \bibinfo{author}{\bibfnamefont{S.}~\bibnamefont{Lizzo}},
  \bibinfo{author}{\bibfnamefont{W.}~\bibnamefont{Smit}}, \bibnamefont{and}
  \bibinfo{author}{\bibfnamefont{G.}~\bibnamefont{Blasse}},
  \bibinfo{journal}{J. Electrochem. Soc.} \textbf{\bibinfo{volume}{140}},
  \bibinfo{pages}{216} (\bibinfo{year}{1993}).

\bibitem[{BAR()}]{BARANDIARAN:14}
 \bibinfo{note}{{Z.} Barandiar\'{a}n and L. Seijo, "Intervalence Charge
  Transfer Luminescence: Interpaly Between Anomalous and 5d-4f emissions in
  Yb-doped florite-type crystals," J. Chem. Phys., accepted.}

\bibitem[{\citenamefont{Bessi\`ere et~al.}(2006)\citenamefont{Bessi\`ere,
  Dorenbos, van Eijk, Kr\"amer, G\"udel, and Galtayries}}]{BESSIERE:06}
\bibinfo{author}{\bibfnamefont{A.}~\bibnamefont{Bessi\`ere}},
  \bibinfo{author}{\bibfnamefont{P.}~\bibnamefont{Dorenbos}},
  \bibinfo{author}{\bibfnamefont{C.}~\bibnamefont{van Eijk}},
  \bibinfo{author}{\bibfnamefont{K.}~\bibnamefont{Kr\"amer}},
  \bibinfo{author}{\bibfnamefont{H.}~\bibnamefont{G\"udel}}, \bibnamefont{and}
  \bibinfo{author}{\bibfnamefont{A.}~\bibnamefont{Galtayries}},
  \bibinfo{journal}{J.\ Lumin.} \textbf{\bibinfo{volume}{117}},
  \bibinfo{pages}{187} (\bibinfo{year}{2006}).

\bibitem[{\citenamefont{Dorenbos et~al.}(2003)\citenamefont{Dorenbos, {van
  Loef}, {van Eijk}, Kr\"amer, and G\"udel}}]{DORENBOS:03:b}
\bibinfo{author}{\bibfnamefont{P.}~\bibnamefont{Dorenbos}},
  \bibinfo{author}{\bibfnamefont{E.}~\bibnamefont{{van Loef}}},
  \bibinfo{author}{\bibfnamefont{C.}~\bibnamefont{{van Eijk}}},
  \bibinfo{author}{\bibfnamefont{K.}~\bibnamefont{Kr\"amer}}, \bibnamefont{and}
  \bibinfo{author}{\bibfnamefont{H.}~\bibnamefont{G\"udel}},
  \bibinfo{journal}{Phys.\ Rev.\ B} \textbf{\bibinfo{volume}{68}},
  \bibinfo{pages}{125108} (\bibinfo{year}{2003}).

\bibitem[{\citenamefont{Bessi\`ere et~al.}(2004)\citenamefont{Bessi\`ere,
  Dorenbos, van Eijk, Pidol, Kr\"amer, and G\"udel}}]{BESSIERE:04}
\bibinfo{author}{\bibfnamefont{A.}~\bibnamefont{Bessi\`ere}},
  \bibinfo{author}{\bibfnamefont{P.}~\bibnamefont{Dorenbos}},
  \bibinfo{author}{\bibfnamefont{C.}~\bibnamefont{van Eijk}},
  \bibinfo{author}{\bibfnamefont{L.}~\bibnamefont{Pidol}},
  \bibinfo{author}{\bibfnamefont{K.}~\bibnamefont{Kr\"amer}}, \bibnamefont{and}
  \bibinfo{author}{\bibfnamefont{H.}~\bibnamefont{G\"udel}},
  \bibinfo{journal}{J.\ Phys.:\ Condens. \ Matter}
  \textbf{\bibinfo{volume}{16}}, \bibinfo{pages}{1887} (\bibinfo{year}{2004}).

\bibitem[{\citenamefont{Duan et~al.}(2009)\citenamefont{Duan, Tanner,
  Meijerink, and Babin}}]{DUAN:09}
\bibinfo{author}{\bibfnamefont{C.-K.} \bibnamefont{Duan}},
  \bibinfo{author}{\bibfnamefont{P.~A.} \bibnamefont{Tanner}},
  \bibinfo{author}{\bibfnamefont{A.}~\bibnamefont{Meijerink}},
  \bibnamefont{and} \bibinfo{author}{\bibfnamefont{V.}~\bibnamefont{Babin}},
  \bibinfo{journal}{J.\ Phys.:\ Condens. \ Matter}
  \textbf{\bibinfo{volume}{21}}, \bibinfo{pages}{395501}
  (\bibinfo{year}{2009}).

\bibitem[{\citenamefont{Duan et~al.}(2011)\citenamefont{Duan, Tanner, Makhov,
  and Khaidukov}}]{DUAN:11}
\bibinfo{author}{\bibfnamefont{C.-K.} \bibnamefont{Duan}},
  \bibinfo{author}{\bibfnamefont{P.~A.} \bibnamefont{Tanner}},
  \bibinfo{author}{\bibfnamefont{V.}~\bibnamefont{Makhov}}, \bibnamefont{and}
  \bibinfo{author}{\bibfnamefont{N.}~\bibnamefont{Khaidukov}},
  \bibinfo{journal}{J.\ Phys.\ Chem.\ A} \textbf{\bibinfo{volume}{115}},
  \bibinfo{pages}{8870} (\bibinfo{year}{2011}).

\bibitem[{\citenamefont{McWeeny}(1959)}]{MCWEENY:59}
\bibinfo{author}{\bibfnamefont{R.}~\bibnamefont{McWeeny}},
  \bibinfo{journal}{Proc. R. Soc. Lond. A} \textbf{\bibinfo{volume}{253}},
  \bibinfo{pages}{242} (\bibinfo{year}{1959}).

\bibitem[{\citenamefont{McWeeny}(1989)}]{McWeeny89}
\bibinfo{author}{\bibfnamefont{R.}~\bibnamefont{McWeeny}},
  \emph{\bibinfo{title}{Methods of molecular quantum mechanics}}
  (\bibinfo{publisher}{{Academic Press}}, \bibinfo{address}{{London}},
  \bibinfo{year}{1989}), \bibinfo{edition}{2nd} ed.

\bibitem[{\citenamefont{Karlstr\"{o}m et~al.}(2003)\citenamefont{Karlstr\"{o}m,
  Lindh, Malmqvist, Roos, Ryde, Veryazov, Widmark, Cossi, Schimmelpfennig,
  Neogrady et~al.}}]{MOLCAS}
\bibinfo{author}{\bibfnamefont{G.}~\bibnamefont{Karlstr\"{o}m}},
  \bibinfo{author}{\bibfnamefont{R.}~\bibnamefont{Lindh}},
  \bibinfo{author}{\bibfnamefont{P.~A.} \bibnamefont{Malmqvist}},
  \bibinfo{author}{\bibfnamefont{B.~O.} \bibnamefont{Roos}},
  \bibinfo{author}{\bibfnamefont{U.}~\bibnamefont{Ryde}},
  \bibinfo{author}{\bibfnamefont{V.}~\bibnamefont{Veryazov}},
  \bibinfo{author}{\bibfnamefont{P.~O.} \bibnamefont{Widmark}},
  \bibinfo{author}{\bibfnamefont{M.}~\bibnamefont{Cossi}},
  \bibinfo{author}{\bibfnamefont{B.}~\bibnamefont{Schimmelpfennig}},
  \bibinfo{author}{\bibfnamefont{P.}~\bibnamefont{Neogrady}},
  \bibnamefont{et~al.}, \bibinfo{journal}{Comput.\ Mater.\ Sci.}
  \textbf{\bibinfo{volume}{28}}, \bibinfo{pages}{222} (\bibinfo{year}{2003}).

\bibitem[{\citenamefont{Douglas and Kroll}(1974)}]{DOUGLAS:74}
\bibinfo{author}{\bibfnamefont{M.}~\bibnamefont{Douglas}} \bibnamefont{and}
  \bibinfo{author}{\bibfnamefont{N.~M.} \bibnamefont{Kroll}},
  \bibinfo{journal}{Ann. Phys. (N.Y.)} \textbf{\bibinfo{volume}{82}},
  \bibinfo{pages}{89} (\bibinfo{year}{1974}).

\bibitem[{\citenamefont{Hess}(1986)}]{HESS:86}
\bibinfo{author}{\bibfnamefont{B.~A.} \bibnamefont{Hess}},
  \bibinfo{journal}{Phys.\ Rev.\ A} \textbf{\bibinfo{volume}{33}},
  \bibinfo{pages}{3742} (\bibinfo{year}{1986}).

\bibitem[{\citenamefont{Roos et~al.}(1980)\citenamefont{Roos, Taylor, and
  Siegbahn}}]{ROOS:80}
\bibinfo{author}{\bibfnamefont{B.~O.} \bibnamefont{Roos}},
  \bibinfo{author}{\bibfnamefont{P.~R.} \bibnamefont{Taylor}},
  \bibnamefont{and} \bibinfo{author}{\bibfnamefont{P.~E.~M.}
  \bibnamefont{Siegbahn}}, \bibinfo{journal}{Chem.\ Phys.}
  \textbf{\bibinfo{volume}{48}}, \bibinfo{pages}{157} (\bibinfo{year}{1980}).

\bibitem[{\citenamefont{Siegbahn et~al.}(1980)\citenamefont{Siegbahn, Heiberg,
  Roos, and Levy}}]{SIEGBAHN:80}
\bibinfo{author}{\bibfnamefont{P.~E.~M.} \bibnamefont{Siegbahn}},
  \bibinfo{author}{\bibfnamefont{A.}~\bibnamefont{Heiberg}},
  \bibinfo{author}{\bibfnamefont{B.~O.} \bibnamefont{Roos}}, \bibnamefont{and}
  \bibinfo{author}{\bibfnamefont{B.}~\bibnamefont{Levy}},
  \bibinfo{journal}{Phys. Scr.} \textbf{\bibinfo{volume}{21}},
  \bibinfo{pages}{323} (\bibinfo{year}{1980}).

\bibitem[{\citenamefont{Siegbahn et~al.}(1981)\citenamefont{Siegbahn, Heiberg,
  Alml{\"o}f, and Roos}}]{SIEGBAHN:81}
\bibinfo{author}{\bibfnamefont{P.~E.~M.} \bibnamefont{Siegbahn}},
  \bibinfo{author}{\bibfnamefont{A.}~\bibnamefont{Heiberg}},
  \bibinfo{author}{\bibfnamefont{J.}~\bibnamefont{Alml{\"o}f}},
  \bibnamefont{and} \bibinfo{author}{\bibfnamefont{B.~O.} \bibnamefont{Roos}},
  \bibinfo{journal}{J.\ Chem.\ Phys.} \textbf{\bibinfo{volume}{74}},
  \bibinfo{pages}{2384} (\bibinfo{year}{1981}).

\bibitem[{\citenamefont{Andersson et~al.}(1990)\citenamefont{Andersson,
  Malmqvist, Roos, Sadlej, and Wolinski}}]{ANDERSSON:90}
\bibinfo{author}{\bibfnamefont{K.}~\bibnamefont{Andersson}},
  \bibinfo{author}{\bibfnamefont{P.-A.} \bibnamefont{Malmqvist}},
  \bibinfo{author}{\bibfnamefont{B.~O.} \bibnamefont{Roos}},
  \bibinfo{author}{\bibfnamefont{A.~J.} \bibnamefont{Sadlej}},
  \bibnamefont{and} \bibinfo{author}{\bibfnamefont{K.}~\bibnamefont{Wolinski}},
  \bibinfo{journal}{J.\ Phys.\ Chem.} \textbf{\bibinfo{volume}{94}},
  \bibinfo{pages}{5483} (\bibinfo{year}{1990}).

\bibitem[{\citenamefont{Andersson et~al.}(1992)\citenamefont{Andersson,
  Malmqvist, and Roos}}]{ANDERSSON:92}
\bibinfo{author}{\bibfnamefont{K.}~\bibnamefont{Andersson}},
  \bibinfo{author}{\bibfnamefont{P.-A.} \bibnamefont{Malmqvist}},
  \bibnamefont{and} \bibinfo{author}{\bibfnamefont{B.~O.} \bibnamefont{Roos}},
  \bibinfo{journal}{J.\ Chem.\ Phys.} \textbf{\bibinfo{volume}{96}},
  \bibinfo{pages}{1218} (\bibinfo{year}{1992}).

\bibitem[{\citenamefont{Zaitsevskii and Malrieu}(1995)}]{ZAITSEVSKII:95}
\bibinfo{author}{\bibfnamefont{A.}~\bibnamefont{Zaitsevskii}} \bibnamefont{and}
  \bibinfo{author}{\bibfnamefont{J.-P.} \bibnamefont{Malrieu}},
  \bibinfo{journal}{Chem.\ Phys.\ Lett.} \textbf{\bibinfo{volume}{233}},
  \bibinfo{pages}{597} (\bibinfo{year}{1995}).

\bibitem[{\citenamefont{Finley et~al.}(1998)\citenamefont{Finley, Malmqvist,
  Roos, and Serrano-Andr\'es}}]{FINLEY:98}
\bibinfo{author}{\bibfnamefont{J.}~\bibnamefont{Finley}},
  \bibinfo{author}{\bibfnamefont{P.-A.} \bibnamefont{Malmqvist}},
  \bibinfo{author}{\bibfnamefont{B.~O.} \bibnamefont{Roos}}, \bibnamefont{and}
  \bibinfo{author}{\bibfnamefont{L.}~\bibnamefont{Serrano-Andr\'es}},
  \bibinfo{journal}{Chem.\ Phys.\ Lett.} \textbf{\bibinfo{volume}{288}},
  \bibinfo{pages}{299} (\bibinfo{year}{1998}).

\bibitem[{\citenamefont{{G. Ghigo, B. O. Roos,} and {P.-\AA.
  Malmqvist}}(2004)}]{GHIGO:04}
\bibinfo{author}{\bibnamefont{{G. Ghigo, B. O. Roos,}}} \bibnamefont{and}
  \bibinfo{author}{\bibnamefont{{P.-\AA. Malmqvist}}}, \bibinfo{journal}{Chem.\
  Phys.\ Lett.} \textbf{\bibinfo{volume}{396}}, \bibinfo{pages}{142}
  (\bibinfo{year}{2004}).

\bibitem[{\citenamefont{Hess et~al.}(1996)\citenamefont{Hess, Marian, Wahlgren,
  and Gropen}}]{HESS:96}
\bibinfo{author}{\bibfnamefont{B.~A.} \bibnamefont{Hess}},
  \bibinfo{author}{\bibfnamefont{C.~M.} \bibnamefont{Marian}},
  \bibinfo{author}{\bibfnamefont{U.}~\bibnamefont{Wahlgren}}, \bibnamefont{and}
  \bibinfo{author}{\bibfnamefont{O.}~\bibnamefont{Gropen}},
  \bibinfo{journal}{Chem.\ Phys.\ Lett.} \textbf{\bibinfo{volume}{251}},
  \bibinfo{pages}{365} (\bibinfo{year}{1996}).

\bibitem[{\citenamefont{Llusar et~al.}(1996)\citenamefont{Llusar, Casarrubios,
  Barandiar\'{a}n, and Seijo}}]{LLUSAR:96}
\bibinfo{author}{\bibfnamefont{R.}~\bibnamefont{Llusar}},
  \bibinfo{author}{\bibfnamefont{M.}~\bibnamefont{Casarrubios}},
  \bibinfo{author}{\bibfnamefont{Z.}~\bibnamefont{Barandiar\'{a}n}},
  \bibnamefont{and} \bibinfo{author}{\bibfnamefont{L.}~\bibnamefont{Seijo}},
  \bibinfo{journal}{J.\ Chem.\ Phys.} \textbf{\bibinfo{volume}{105}},
  \bibinfo{pages}{5321} (\bibinfo{year}{1996}).

\bibitem[{\citenamefont{Malmqvist et~al.}(2002)\citenamefont{Malmqvist, Roos,
  and Schimmelpfennig}}]{MALMQVIST:02}
\bibinfo{author}{\bibfnamefont{P.~A.} \bibnamefont{Malmqvist}},
  \bibinfo{author}{\bibfnamefont{B.~O.} \bibnamefont{Roos}}, \bibnamefont{and}
  \bibinfo{author}{\bibfnamefont{B.}~\bibnamefont{Schimmelpfennig}},
  \bibinfo{journal}{Chem.\ Phys.\ Lett.} \textbf{\bibinfo{volume}{357}},
  \bibinfo{pages}{230} (\bibinfo{year}{2002}).

\bibitem[{\citenamefont{Paulovic et~al.}(2003)\citenamefont{Paulovic, Nakajima,
  Hirao, Lindh, , and Malmqvist}}]{PAULOVIC:03}
\bibinfo{author}{\bibfnamefont{J.}~\bibnamefont{Paulovic}},
  \bibinfo{author}{\bibfnamefont{T.}~\bibnamefont{Nakajima}},
  \bibinfo{author}{\bibfnamefont{K.}~\bibnamefont{Hirao}},
  \bibinfo{author}{\bibfnamefont{R.}~\bibnamefont{Lindh}}, , \bibnamefont{and}
  \bibinfo{author}{\bibfnamefont{P.-A.} \bibnamefont{Malmqvist}},
  \bibinfo{journal}{J.\ Chem.\ Phys.} \textbf{\bibinfo{volume}{119}},
  \bibinfo{pages}{798} (\bibinfo{year}{2003}).

\bibitem[{\citenamefont{Roos et~al.}(2008)\citenamefont{Roos, Lindh, Malmqvist,
  Veryazov, and Widmark}}]{ROOS:08}
\bibinfo{author}{\bibfnamefont{B.~O.} \bibnamefont{Roos}},
  \bibinfo{author}{\bibfnamefont{R.}~\bibnamefont{Lindh}},
  \bibinfo{author}{\bibfnamefont{P.~A.} \bibnamefont{Malmqvist}},
  \bibinfo{author}{\bibfnamefont{V.}~\bibnamefont{Veryazov}}, \bibnamefont{and}
  \bibinfo{author}{\bibfnamefont{P.~O.} \bibnamefont{Widmark}},
  \bibinfo{journal}{J.\ Chem.\ Phys.} \textbf{\bibinfo{volume}{112}},
  \bibinfo{pages}{11431} (\bibinfo{year}{2008}).

\bibitem[{\citenamefont{Roos et~al.}(2005)\citenamefont{Roos, Lindh, Malmqvist,
  Veryazov, and Widmark}}]{ROOS:05}
\bibinfo{author}{\bibfnamefont{B.~O.} \bibnamefont{Roos}},
  \bibinfo{author}{\bibfnamefont{R.}~\bibnamefont{Lindh}},
  \bibinfo{author}{\bibfnamefont{P.~A.} \bibnamefont{Malmqvist}},
  \bibinfo{author}{\bibfnamefont{V.}~\bibnamefont{Veryazov}}, \bibnamefont{and}
  \bibinfo{author}{\bibfnamefont{P.~O.} \bibnamefont{Widmark}},
  \bibinfo{journal}{J.\ Phys.\ Chem.\ A} \textbf{\bibinfo{volume}{108}},
  \bibinfo{pages}{2851} (\bibinfo{year}{2005}).

\bibitem[{\citenamefont{Roos et~al.}(2003)\citenamefont{Roos, Veryazov, and
  Widmark}}]{ROOS:03}
\bibinfo{author}{\bibfnamefont{B.~O.} \bibnamefont{Roos}},
  \bibinfo{author}{\bibfnamefont{V.}~\bibnamefont{Veryazov}}, \bibnamefont{and}
  \bibinfo{author}{\bibfnamefont{P.~O.} \bibnamefont{Widmark}},
  \bibinfo{journal}{Theor. Chim. Acta} \textbf{\bibinfo{volume}{111}},
  \bibinfo{pages}{345} (\bibinfo{year}{2003}).

\bibitem[{\citenamefont{Huzinaga et~al.}(1987)\citenamefont{Huzinaga, Seijo,
  Barandiar\'an, and Klobukowski}}]{HUZINAGA:87}
\bibinfo{author}{\bibfnamefont{S.}~\bibnamefont{Huzinaga}},
  \bibinfo{author}{\bibfnamefont{L.}~\bibnamefont{Seijo}},
  \bibinfo{author}{\bibfnamefont{Z.}~\bibnamefont{Barandiar\'an}},
  \bibnamefont{and}
  \bibinfo{author}{\bibfnamefont{M.}~\bibnamefont{Klobukowski}},
  \bibinfo{journal}{J.\ Chem.\ Phys.} \textbf{\bibinfo{volume}{86}},
  \bibinfo{pages}{2132} (\bibinfo{year}{1987}).

\bibitem[{\citenamefont{Barandiar\'{a}n and Seijo}(1988)}]{BARANDIARAN:88}
\bibinfo{author}{\bibfnamefont{Z.}~\bibnamefont{Barandiar\'{a}n}}
  \bibnamefont{and} \bibinfo{author}{\bibfnamefont{L.}~\bibnamefont{Seijo}},
  \bibinfo{journal}{J.\ Chem.\ Phys.} \textbf{\bibinfo{volume}{89}},
  \bibinfo{pages}{5739} (\bibinfo{year}{1988}).

\bibitem[{\citenamefont{Gell\'{e} and Lepetit}(2008)}]{GELLE:08}
\bibinfo{author}{\bibfnamefont{A.}~\bibnamefont{Gell\'{e}}} \bibnamefont{and}
  \bibinfo{author}{\bibfnamefont{M.-B.} \bibnamefont{Lepetit}},
  \bibinfo{journal}{J.\ Chem.\ Phys.} \textbf{\bibinfo{volume}{128}},
  \bibinfo{pages}{244716} (\bibinfo{year}{2008}).

\bibitem[{\citenamefont{Ewald}(1921)}]{EWALD:21}
\bibinfo{author}{\bibfnamefont{P.~P.} \bibnamefont{Ewald}},
  \bibinfo{journal}{Ann. Phys.} \textbf{\bibinfo{volume}{369}},
  \bibinfo{pages}{253} (\bibinfo{year}{1921}).

\bibitem[{\citenamefont{Meyer et~al.}(1986)\citenamefont{Meyer, Hwu, and
  Corbett}}]{MEYER:86}
\bibinfo{author}{\bibfnamefont{G.}~\bibnamefont{Meyer}},
  \bibinfo{author}{\bibfnamefont{S.~J.} \bibnamefont{Hwu}}, \bibnamefont{and}
  \bibinfo{author}{\bibfnamefont{J.~D.} \bibnamefont{Corbett}},
  \bibinfo{journal}{Z. Anorg. Allg. Chem.} \textbf{\bibinfo{volume}{535}},
  \bibinfo{pages}{208} (\bibinfo{year}{1986}).

\bibitem[{\citenamefont{Seijo and Barandiar\'{a}n}(1991)}]{SEIJO:91}
\bibinfo{author}{\bibfnamefont{L.}~\bibnamefont{Seijo}} \bibnamefont{and}
  \bibinfo{author}{\bibfnamefont{Z.}~\bibnamefont{Barandiar\'{a}n}},
  \bibinfo{journal}{J.\ Chem.\ Phys.} \textbf{\bibinfo{volume}{94}},
  \bibinfo{pages}{8158} (\bibinfo{year}{1991}).

\bibitem[{AIM()}]{AIMP-EMBEDDING-DATA}
\bibinfo{note}{Detailed em\-bed\-ding AIMP data li\-bra\-ries in elec\-tro\-nic
  format are avail\-able from the au\-thors upon re\-quest or di\-rectly at the
  ad\-dress http:\-//www.uam.es/\-quimica/\-aimp/\-Data/\-AIMPLibs.html. See
  also Ref.~\onlinecite{MOLCAS}.}

\bibitem[{\citenamefont{Bersuker}(1984)}]{BERSUKER:84}
\bibinfo{author}{\bibfnamefont{I.~B.} \bibnamefont{Bersuker}},
  \emph{\bibinfo{title}{The Jahn-Teller Effect and Vibronic Interactions in
  Modern Chemistry}} (\bibinfo{publisher}{Plenum Press}, \bibinfo{address}{New
  York and London}, \bibinfo{year}{1984}).

\bibitem[{\citenamefont{Tanner et~al.}(2003)\citenamefont{Tanner, Mak,
  Edelstein, Murdoch, Liu, Huang, Seijo, , and Barandiar\'an}}]{TANNER:03}
\bibinfo{author}{\bibfnamefont{P.~A.} \bibnamefont{Tanner}},
  \bibinfo{author}{\bibfnamefont{C.~S.~K.} \bibnamefont{Mak}},
  \bibinfo{author}{\bibfnamefont{N.~M.} \bibnamefont{Edelstein}},
  \bibinfo{author}{\bibfnamefont{K.~M.} \bibnamefont{Murdoch}},
  \bibinfo{author}{\bibfnamefont{G.}~\bibnamefont{Liu}},
  \bibinfo{author}{\bibfnamefont{J.}~\bibnamefont{Huang}},
  \bibinfo{author}{\bibfnamefont{L.}~\bibnamefont{Seijo}}, , \bibnamefont{and}
  \bibinfo{author}{\bibfnamefont{Z.}~\bibnamefont{Barandiar\'an}},
  \bibinfo{journal}{J.\ Amer.\ Chem.\ Soc.} \textbf{\bibinfo{volume}{125}},
  \bibinfo{pages}{13225} (\bibinfo{year}{2003}).

\bibitem[{\citenamefont{Ordej\'on et~al.}(2007)\citenamefont{Ordej\'on, Seijo,
  and Barandiar\'an}}]{ORDEJON:07}
\bibinfo{author}{\bibfnamefont{B.}~\bibnamefont{Ordej\'on}},
  \bibinfo{author}{\bibfnamefont{L.}~\bibnamefont{Seijo}}, \bibnamefont{and}
  \bibinfo{author}{\bibfnamefont{Z.}~\bibnamefont{Barandiar\'an}},
  \bibinfo{journal}{J.\ Chem.\ Phys.} \textbf{\bibinfo{volume}{126}},
  \bibinfo{pages}{194712} (\bibinfo{year}{2007}).

\bibitem[{\citenamefont{S\'anchez-Sanz
  et~al.}(2010)\citenamefont{S\'anchez-Sanz, Seijo, and
  Barandiar\'an}}]{SANCHEZ-SANZ:10:a}
\bibinfo{author}{\bibfnamefont{G.}~\bibnamefont{S\'anchez-Sanz}},
  \bibinfo{author}{\bibfnamefont{L.}~\bibnamefont{Seijo}}, \bibnamefont{and}
  \bibinfo{author}{\bibfnamefont{Z.}~\bibnamefont{Barandiar\'an}},
  \bibinfo{journal}{J.\ Chem.\ Phys.} \textbf{\bibinfo{volume}{133}},
  \bibinfo{pages}{114509} (\bibinfo{year}{2010}).

\bibitem[{\citenamefont{Heller}(1975)}]{HELLER:75}
\bibinfo{author}{\bibfnamefont{E.~J.} \bibnamefont{Heller}},
  \bibinfo{journal}{J.\ Chem.\ Phys.} \textbf{\bibinfo{volume}{62}},
  \bibinfo{pages}{1544} (\bibinfo{year}{1975}).

\bibitem[{\citenamefont{Heller}(1981)}]{HELLER:81}
\bibinfo{author}{\bibfnamefont{E.~J.} \bibnamefont{Heller}},
  \bibinfo{journal}{Acc.\ Chem.\ Res.} \textbf{\bibinfo{volume}{14}},
  \bibinfo{pages}{368} (\bibinfo{year}{1981}).

\bibitem[{\citenamefont{Zink and Shin}(1991)}]{ZINK:91:b}
\bibinfo{author}{\bibfnamefont{J.~I.} \bibnamefont{Zink}} \bibnamefont{and}
  \bibinfo{author}{\bibfnamefont{K.~S.} \bibnamefont{Shin}}, in
  \emph{\bibinfo{booktitle}{Advances in Photochemistry}}
  (\bibinfo{publisher}{Wiley}, \bibinfo{address}{New York},
  \bibinfo{year}{1991}), vol.~\bibinfo{volume}{16}, pp.
  \bibinfo{pages}{119--214}.

\bibitem[{\citenamefont{McClure and P\'edrini}(1985)}]{MCCLURE:85}
\bibinfo{author}{\bibfnamefont{D.~S.} \bibnamefont{McClure}} \bibnamefont{and}
  \bibinfo{author}{\bibfnamefont{C.}~\bibnamefont{P\'edrini}},
  \bibinfo{journal}{Phys.\ Rev.\ B} \textbf{\bibinfo{volume}{32}},
  \bibinfo{pages}{8465} (\bibinfo{year}{1985}).

\bibitem[{\citenamefont{{B. Moine, B. Courtois} and
  P\'edrini}(1989)}]{MOINE:89}
\bibinfo{author}{\bibnamefont{{B. Moine, B. Courtois}}} \bibnamefont{and}
  \bibinfo{author}{\bibfnamefont{C.}~\bibnamefont{P\'edrini}},
  \bibinfo{journal}{J. Phys. France} \textbf{\bibinfo{volume}{50}},
  \bibinfo{pages}{2105} (\bibinfo{year}{1989}).

\bibitem[{\citenamefont{Dorenbos}(2003)}]{DORENBOS:03:a}
\bibinfo{author}{\bibfnamefont{P.}~\bibnamefont{Dorenbos}},
  \bibinfo{journal}{J.\ Phys.:\ Condens. \ Matter}
  \textbf{\bibinfo{volume}{15}}, \bibinfo{pages}{2645} (\bibinfo{year}{2003}).

\bibitem[{\citenamefont{Gracia et~al.}(2008)\citenamefont{Gracia, Seijo,
  Barandiar\'an, Curulla, Niemansverdriet, and van Gennip}}]{GRACIA:08}
\bibinfo{author}{\bibfnamefont{J.}~\bibnamefont{Gracia}},
  \bibinfo{author}{\bibfnamefont{L.}~\bibnamefont{Seijo}},
  \bibinfo{author}{\bibfnamefont{Z.}~\bibnamefont{Barandiar\'an}},
  \bibinfo{author}{\bibfnamefont{D.}~\bibnamefont{Curulla}},
  \bibinfo{author}{\bibfnamefont{H.}~\bibnamefont{Niemansverdriet}},
  \bibnamefont{and} \bibinfo{author}{\bibfnamefont{W.}~\bibnamefont{van
  Gennip}}, \bibinfo{journal}{J.\ Lumin.} \textbf{\bibinfo{volume}{128}},
  \bibinfo{pages}{1248} (\bibinfo{year}{2008}).

\end{thebibliography}
\bibliographystyle{apsrev}

\begin{widetext}
\iftoggle{usualpreprint}{
    \input{tables.tex}
}{}
\iftoggle{journal-like-bigfigures}{
    \input{tables.tex}
}{}
%
\iftoggle{journal-like-bigfigures}{
    \def\escalafig{0.6}\clearpage 
    \begin{figure}[ht]
      \input{FIG-\FIGI-fig.tex}
      \input{FIG-\FIGI-cap.tex}
    \end{figure}
    \def\escalafig{0.7}\clearpage 
    \begin{figure}[ht]
      \input{FIG-\FIGII-fig.tex}
      \input{FIG-\FIGII-cap.tex}
    \end{figure}
    \def\escalafig{0.6}\clearpage 
    \begin{figure}[ht]
      \input{FIG-\FIGIII-fig.tex}
      \input{FIG-\FIGIII-cap.tex}
    \end{figure}
    \def\escalafig{0.7}\clearpage 
    \begin{figure}[ht]
      \input{FIG-\FIGIV-fig.tex}
      \input{FIG-\FIGIV-cap.tex}
    \end{figure}
    \def\escalafig{1.0}\clearpage 
    \begin{figure}[ht]
      \input{FIG-\FIGV-fig.tex}
      \input{FIG-\FIGV-cap.tex}
    \end{figure}
    \def\escalafig{0.6}\clearpage 
    \begin{figure}[ht]
      \input{FIG-\FIGVI-fig.tex}
      \input{FIG-\FIGVI-cap.tex}
    \end{figure}
    \def\escalafig{0.6}\clearpage 
    \begin{figure}[ht]
      \input{FIG-\FIGVII-fig.tex}
      \input{FIG-\FIGVII-cap.tex}
    \end{figure}
}{}
\iftoggle{usualpreprint}{
    \clearpage \section*{Figure captions}
    \begin{figure}[ht]
      \input{FIG-\FIGI-cap.tex}
    \end{figure}
    \begin{figure}[ht]
      \input{FIG-\FIGII-cap.tex}
    \end{figure}
    \begin{figure}[ht]
      \input{FIG-\FIGIII-cap.tex}
    \end{figure}
    \begin{figure}[ht]
      \input{FIG-\FIGIV-cap.tex}
    \end{figure}
    \begin{figure}[ht]
      \input{FIG-\FIGV-cap.tex}
    \end{figure}
    \begin{figure}[ht]
      \input{FIG-\FIGVI-cap.tex}
    \end{figure}
    \begin{figure}[ht]
      \input{FIG-\FIGVII-cap.tex}
    \end{figure}
    \vfill\mbox{}
    \def\escalafig{0.6}
    \clearpage\begin{center}
    \input{FIG-\FIGI-fig.tex}
    \vfill 
    Figure~\ref{FIG:\FIGI}. 
    \end{center}\thispagestyle{empty}
    \def\escalafig{0.6}
    \clearpage\begin{center}
    \input{FIG-\FIGII-fig.tex}
    \vfill 
    Figure~\ref{FIG:\FIGII}. 
    \end{center}\thispagestyle{empty}
    \def\escalafig{0.6}
    \clearpage\begin{center}
    \input{FIG-\FIGIII-fig.tex}
    \vfill 
    Figure~\ref{FIG:\FIGIII}. 
    \end{center}\thispagestyle{empty}
    \def\escalafig{0.9}
    \clearpage\begin{center}
    \input{FIG-\FIGIV-fig.tex}
    \vfill 
    Figure~\ref{FIG:\FIGIV}. 
    \end{center}\thispagestyle{empty}
    \def\escalafig{0.9}
    \clearpage\begin{center}
    \input{FIG-\FIGV-fig.tex}
    \vfill 
    Figure~\ref{FIG:\FIGV}. 
    \end{center}\thispagestyle{empty}
    \def\escalafig{0.9}
    \clearpage\begin{center}
    \input{FIG-\FIGV-fig.tex}
    \vfill 
    Figure~\ref{FIG:\FIGVI}. 
    \end{center}\thispagestyle{empty}
    \def\escalafig{0.9}
    \clearpage\begin{center}
    \input{FIG-\FIGV-fig.tex}
    \vfill 
    Figure~\ref{FIG:\FIGVII}. 
    \end{center}\thispagestyle{empty}
}{}


 \clearpage
 \section{\label{SEC:Add}Additional material}
\clearpage
\begin{table}[h]
\caption{
         Computed absorption and emisson oscillator strenthgs
         of the \CeIII\ active center in Ce-doped \HostCsLiLuCl. 
              }
\label{TAB:oscil_strengths}
\begin{ruledtabular}
\begin{tabular}{cccccc}
 Configuration & State & $f_{abs}(1\Gamma_{7u}\rightarrow)$ 
                       & $f_{em}(1\Gamma_{8g}\rightarrow)$ 
                       & $f_{em}(2\Gamma_{8g}\rightarrow)$
                       & $f_{em}(1\Gamma_{6g}\rightarrow)$ \\
$4f^1$  \\
  & $1\Gamma_{7u}$($^2A_{2u}$)  &  & $1.08\times 10^{-2}$ & $2.14\times 10^{-3}$ & 0    \\
  & $1\Gamma_{8u}$($^2T_{2u}$)  &  & $1.67\times 10^{-2}$ & $2.74\times 10^{-2}$ & $3.29\times 10^{-5}$ \\
  & $2\Gamma_{7u}$($^2T_{2u}$)  &  & $1.62\times 10^{-4}$ & $3.31\times 10^{-3}$ & 0     \\
  & $2\Gamma_{8u}$($^2T_{1u}$)  &  & $5.79\times 10^{-3}$ & $2.67\times 10^{-2}$ & $2.60\times 10^{-5}$ \\
  & $1\Gamma_{6u}$($^2T_{1u}$)  &  & $5.33\times 10^{-3}$ & $2.23\times 10^{-2}$ & $2.86\times 10^{-5}$ \\
$5dt_{2g}^1$   \\                                      
  & $1\Gamma_{8g}$($^2T_{2g}$)  & $2.15\times 10^{-2}$ &  &     \\
  & $1\Gamma_{7g}$($^2T_{2g}$)  & $1.54\times 10^{-4}$ &  &     \\
$5de_{g}^1$   \\                                      
  & $2\Gamma_{8g}$($^2E_{g}$)   & $4.28\times 10^{-3}$ \\
ITE-$a_{1g}^1$   \\                                      
  & $1\Gamma_{6g}$($^2A_{1g}$)  & 0  \\
\end{tabular}
\end{ruledtabular}

\end{table}
\clearpage
\begin{table}[h]
\caption{
         Absorption and emission horizontal offsets of the \CeIII--\CeIV\ pair in \HostCsLiLuCl,
         $\Delta Q = Q_{e,f} - Q_{e,i}$,
         in \AA.
         In \CeIII\ regular transitions, $\Delta Q = \sqrt{6} (d_{{\rm Ce-Cl},e,f}-d_{{\rm Ce-Cl},e,i})$.
         In IVCT transitions, Eq.~\ref{EQ:Qet} is used for $Q_{e,f}$ and $Q_{e,i}$,
         using activated complex, and equilibrium Ce-Cl bond lengths of the left and right clusters
         from Table~\ref{TAB:\TABc}.
         These values are approximately equal to $\sqrt{3} (\Delta d_R - \Delta d_L)$,
         $\Delta d_R$ and $\Delta d_L$ being the Ce-Cl bond length changes experienced by 
the \CeIV\ acceptor center (right)
and the \CeIII\ donor center (left) in the IVCT transition.
              }
\label{TAB:CeIII-CeIV-offset}
\begin{ruledtabular}
\begin{tabular}{cccccc}
          && $4f$ absorption & $5dt_{2g}$ emission & $5de_g$ emission & $a_{1g}$ ITE emission \\
           & {Initial state [\CeIII,\CeIV]:} 
           & $[1\Gamma_{7u}$($^2A_{2u}$),$A_{1g}]$ $\rightarrow$
           & $[1\Gamma_{8g}$($^2T_{2g}$),$A_{1g}]$ $\rightarrow$
           & $[2\Gamma_{8g}$($^2E_{g}$),$A_{1g}]$  $\rightarrow$
           & $[1\Gamma_{6g}$($^2A_{1g}$),$A_{1g}]$ $\rightarrow$ \\
 Branch & Final state \\
&                                                    &       &{\bf (2)}&{\bf (4)}&{\bf (6)}\\
  \multicolumn{1}{c}{$A$} 
& \multicolumn{1}{c}{[\CeIII($4f^1$),\CeIV]}         
    & \CeIII\ $4f         \rightarrow 4f$ 
    & \CeIII\ $5dt_{2g}   \rightarrow 4f$ 
    & \CeIII\ $5de_{g}    \rightarrow 4f$ 
    & \CeIII\ ITE-$a_{1g} \rightarrow 4f$ \\ 
  & $\rightarrow$ $[1\Gamma_{7u}(^2A_{2u}),A_{1g}]$  & 0      & +0.103 & --0.022 & +0.257 \\
  & $\rightarrow$ $[1\Gamma_{8u}(^2T_{2u}),A_{1g}]$  & +0.005 & +0.108 & --0.017 & +0.262 \\
  & $\rightarrow$ $[2\Gamma_{7u}(^2T_{2u}),A_{1g}]$  & +0.000 & +0.103 & --0.022 & +0.257 \\
  & $\rightarrow$ $[2\Gamma_{8u}(^2T_{1u}),A_{1g}]$  & +0.005 & +0.108 & --0.017 & +0.262 \\
  & $\rightarrow$ $[1\Gamma_{6u}(^2T_{1u}),A_{1g}]$  & +0.007 & +0.110 & --0.015 & +0.265 \\
&                                                    &{\bf (7)}&      &{\bf (10)} \\ 
  \multicolumn{1}{c}{$A_{\rm et}^\wedge$ } 
& \multicolumn{1}{c}{[\CeIV,\CeIII($4f^1$)]$_{\rm sts}$} 
    &\CeIII $4f$         $\rightarrow$ \CeIV $4f$
    &\CeIII $5dt_{2g}$   $\rightarrow$ \CeIV $4f$
    &\CeIII $5de_g$      $\rightarrow$ \CeIV $4f$ \\
  & $\rightarrow$ $[A_{1g},1\Gamma_{7u}(^2A_{2u})]$  & 0.409 & 0.343 & 0.425 &  \\
  & $\rightarrow$ $[A_{1g},1\Gamma_{8u}(^2T_{2u})]$  & 0.412 & 0.348 & 0.428 &  \\
  & $\rightarrow$ $[A_{1g},2\Gamma_{7u}(^2T_{2u})]$  & 0.409 & 0.343 & 0.425 &  \\
  & $\rightarrow$ $[A_{1g},2\Gamma_{8u}(^2T_{1u})]$  & 0.412 & 0.348 & 0.428 &  \\
  & $\rightarrow$ $[A_{1g},1\Gamma_{6u}(^2T_{1u})]$  & 0.414 & 0.350 & 0.430 &  \\
&                                                  &{\bf (1)}&        &        &        \\
  \multicolumn{1}{c}{$B$ } 
& \multicolumn{1}{c}{[\CeIII($5dt_{2g}^1$),\CeIV]} 
    &\CeIII $4f         \rightarrow 5dt_{2g}$
    &        
    &        
    &        \\
  & $\rightarrow$ $[1\Gamma_{8g}(^2T_{2g}),A_{1g}]$  & --0.103 &  &  &  \\
  & $\rightarrow$ $[1\Gamma_{7g}(^2T_{2g}),A_{1g}]$  & --0.103 &  &  &   \\
&                                                              & {\bf (8)} \\
  \multicolumn{1}{c}{$B_{\rm et}^\wedge$ } 
& \multicolumn{1}{c}{[\CeIV,\CeIII($5dt_{2g}^1$)]$_{\rm sts}$} 
    &\CeIII $4f$         $\rightarrow$ \CeIV $5dt_{2g}$
    &        
    &        
    &       \\
  & $\rightarrow$ $[A_{1g},1\Gamma_{8g}(^2T_{2g})]$  & 0.343 &  &  &   \\
  & $\rightarrow$ $[A_{1g},1\Gamma_{7g}(^2T_{2g})]$  & 0.343 &  &  &   \\
&                                                  &{\bf (3)}&        &        &       \\
  \multicolumn{1}{c}{$C$ } 
& \multicolumn{1}{c}{[\CeIII($5de_{g}^1$),\CeIV]}  
    &\CeIII $4f         \rightarrow 5de_g$ 
    &        
    &        
    &       \\
  & $\rightarrow$ $[2\Gamma_{8g}(^2E_{g}),A_{1g}]$   & +0.022 &  &  &  \\
&                                                              &{\bf (9)} \\
  \multicolumn{1}{c}{$C_{\rm et}^\wedge$ } 
& \multicolumn{1}{c}{[\CeIV,\CeIII($5de_{g}^1$)]$_{\rm sts}$}  
    &\CeIII $4f$         $\rightarrow$ \CeIV $5de_g$
    &        
    &        
    &       \\
  & $\rightarrow$ $[A_{1g},2\Gamma_{8g}(^2E_{g})]$   & 0.425 &  &  &  \\
&                                                      &{\bf (5)}&        &        &       \\
  \multicolumn{1}{c}{$D$ } 
& \multicolumn{1}{c}{[\CeIII(ITE-$a_{1g}\,^1$),\CeIV]} 
    &\CeIII $4f         \rightarrow$ ITE-$a_{1g}$ 
    &        
    &        
    &       \\
  & $\rightarrow$ $[1\Gamma_{6g}(^2A_{1g}),A_{1g}]$  & --0.257 &        &        &  \\
  \multicolumn{1}{c}{$D_{\rm et}^\wedge$ } 
& \multicolumn{1}{c}{[\CeIV,\CeIII(ITE-$a_{1g}\,^1$)]$_{\rm sts}$} 
    &
    &        
    &        
    &       \\
  & $\rightarrow$ $[A_{1g},1\Gamma_{6g}(^2A_{1g})]$  & 0.291 &        &        &  \\
\end{tabular}
\end{ruledtabular}
\end{table}

%
\iftoggle{usualpreprint}{
    \clearpage \section*{Figure captions}
    \begin{figure}[ht]
      \input{FIG-\FIGVII-cap.tex}
    \end{figure}
    \begin{figure}[ht]
      \input{FIG-\FIGVIII-cap.tex}
    \end{figure}
    \vfill\mbox{}
    \def\escalafig{0.7}
    \clearpage\begin{center}
    \input{FIG-\FIGVII-fig.tex}
    \vfill 
    Figure~\ref{FIG:\FIGVII}. 
    \end{center}\thispagestyle{empty}
    \def\escalafig{0.9}
    \clearpage\begin{center}
    \input{FIG-\FIGVIII-fig.tex}
    \vfill 
    Figure~\ref{FIG:\FIGVIII}. 
    \end{center}\thispagestyle{empty}
}{
    \def\escalafig{0.9}\clearpage 
    \begin{figure}[ht]
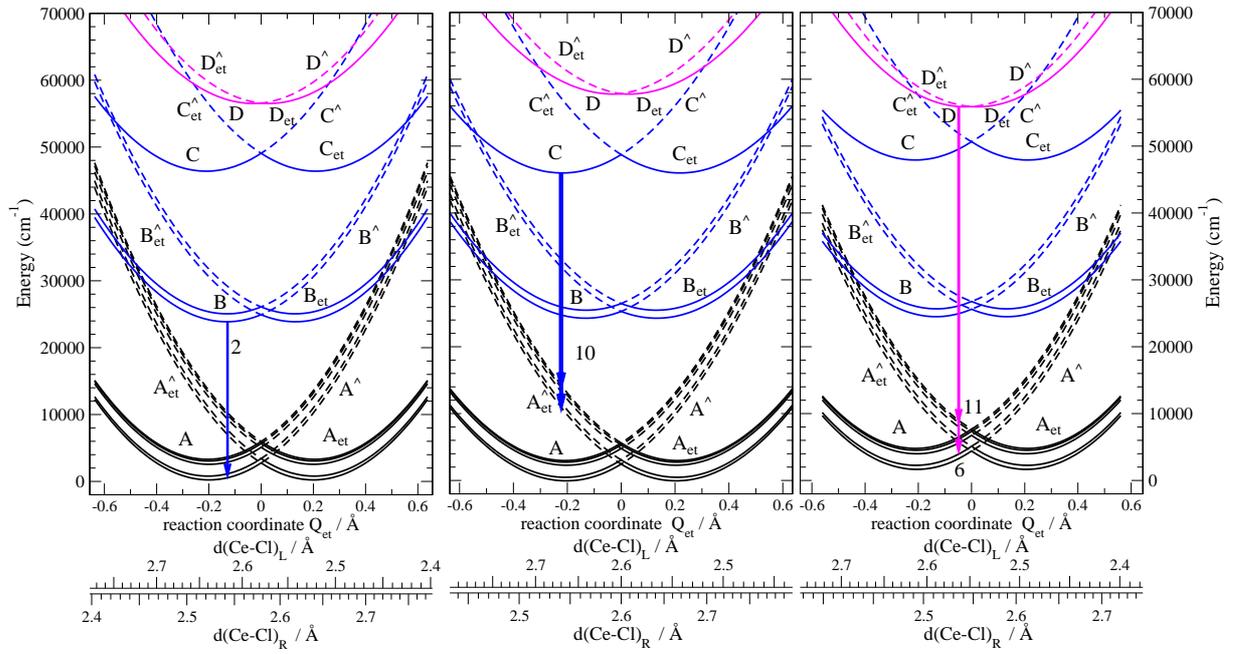

\resizebox{\escalafig\textwidth}{!}{
  \begin{tabular}{ccc}
    \rotatebox{00}{\includegraphics[scale=1,clip]{CeIII-CeIV-61.eps}} &
    \rotatebox{00}{\includegraphics[scale=1,clip]{CeIII-CeIV-81.eps}} &
    \rotatebox{00}{\includegraphics[scale=1,clip]{CeIII-CeIV-91.eps}}
  \end{tabular}
}
\caption{
         Energy diagrams of the \CeIII-\CeIV\ pair:
         Total energies of the \CeClvimiii-\CeClvimii\ cluster pair embedded in \HostCsLiLuCl\
         vs. the electron transfer reaction coordinates 
         of the $5d$ states
         [$1\Gamma_{8g}(t_{2g})$,$A_{1g}$] (left panel) and
         [$2\Gamma_{8g}(e_{g})$,$A_{1g}$] (middle panel),
         and of the impurity-trapped exciton state
         [$1\Gamma_{6g}(a_{1g})$,$A_{1g}$] (right panel).
         The Ce-Cl distances of the corresponding activated complexes are 2.578~\AA, 2.600~\AA, and 2.548~\AA\ respectively.
         Spin-orbit coupling RASSI calculations in the zero electronic coupling limit.
         Energy scale relative to the ground state energy at equilibrium.
         See text for a description of the labels.
}
\label{FIG:CeIII-CeIV-several}
    \end{figure}
    \def\escalafig{0.7}\clearpage 
    \begin{figure}[ht]
\resizebox{\escalafig\textwidth}{!}{
    \rotatebox{-90}{\includegraphics[scale=1,clip]{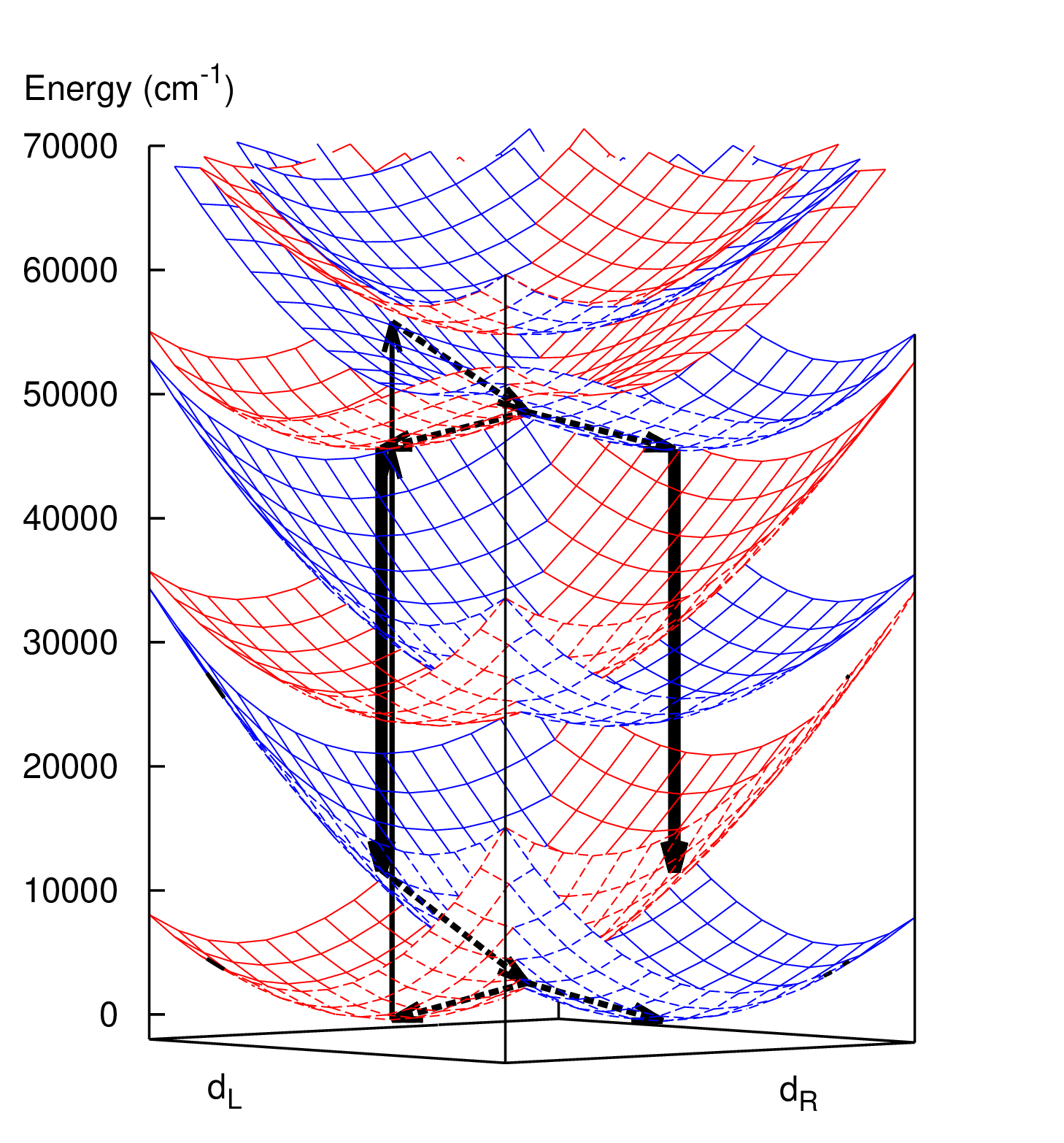}}
}
\caption{
Anomalous emission in Ce-doped \HostCsLiLuCl\ (thick down arrows)
interpreted as a \mbox{\CeIII $5de_g$} $\rightarrow$ \mbox{\CeIV $4f$} IVCT luminescence
taking place between the equilibrium structure of {[\CeIII($5de_{g}^1$),\CeIV]}
and the stressed structure of {[\CeIV,\CeIII($4f^1$)]}.
Excitations of the IVCT luminescence with a \CeIII\ $4f \rightarrow 5de_{g}$ absorption
and a \mbox{\CeIII $4f$} $\rightarrow$ \mbox{\CeIV $5de_g$} IVCT absorption
are shown with thin up arrows followed by dashed arrows.
Nuclei reorganization around the \CeIV\ and \CeIII\ centers after the IVCT luminescence 
is shown with dashed arrows on the lowest energy surfaces.
$d_R$ and $d_L$ stand for the Ce-Cl distances in the left and right Ce centers,
which are, respectively, \CeIII\ and \CeIV\ in the red energy surfaces,
and \CeIV\ and \CeIII\ in the blue surfaces.
}
\label{FIG:ivctl}
    \end{figure}
}

\clearpage 
\end{widetext} 

\end{document}